\documentclass[usenatbib,usegraphicx,useAMS]{mn2e}

\usepackage{times}
\usepackage{amsmath}
\usepackage{amssymb}
\usepackage{bbold}

\newcommand{\mbi}[1]{\mbox{\boldmath$#1$}}

\newcommand{\1}{\mathbb{1}}
\newcommand{\mat}[1]{\mbox{\rm\bf #1}}
\newcommand{\lsim}[1]{\mbox{${\,\hbox{\hbox{$ < $}\kern -0.8em \lower 1.0ex\hbox{$\sim$}}\,}$}}
\newcommand{\gsim}[1]{\mbox{${\,\hbox{\hbox{$ > $}\kern -0.8em \lower 1.0ex\hbox{$\sim$}}\,}$}}

\def\etal{{\it et al.\ }}

\def\beqn{\vspace{2mm}
\begin{eqnarray}} 
\def\eeqn{\vspace{2mm} 
\end{eqnarray}}

 \def\la{\langle}

\begin{document}
\title[Poisson--Lognormal Reconstruction]{Recovering the nonlinear  density field from the galaxy distribution  with a Poisson--Lognormal filter}

\author[Kitaura \etal]{Francisco-Shu Kitaura$^{1,2}$\thanks{E-mail: kitaura@sissa.it, kitaura@mpa-garching.mpg.de}, Jens Jasche$^2$ and  R.~Benton Metcalf$^2$\\
 $^1$  SISSA, Scuola Internazionale Superiore di Studi Avanzati, via Beirut 2-4 34014 Trieste, Italy\\
$^2$ MPA, Max-Planck Institut f\"ur Astrophysik, Karl-Schwarzschildstr.~1, D-85748 Garching, Germany\\
}

\maketitle

\begin{abstract}

We present a general expression for a lognormal filter given an arbitrary nonlinear galaxy bias. We derive this filter as the maximum a posteriori solution  assuming a lognormal prior distribution for the matter field with a given mean field and modeling the observed galaxy distribution by a Poissonian process. We have performed a three--dimensional implementation of this filter  with a very efficient Newton--Krylov inversion scheme. Furthermore, we have tested it with a dark matter N--body simulation assuming a unit galaxy bias relation and compared the results  with previous density field estimators like the inverse weighting scheme and Wiener filtering. 
Our results show good agreement with the underlying dark matter field for overdensities even above $\delta\sim1000$ which exceeds by one order of magnitude the  regime in which the lognormal is expected to be valid. 
The reason is that for our filter the lognormal assumption enters as a prior distribution function, but the maximum a posteriori solution is also conditioned on the data. We find that the lognormal filter is superior to the previous filtering schemes in terms of higher correlation coefficients and smaller Euclidean distances to the underlying matter field. We also show how it is able to recover the positive tail of the matter density field distribution for a unit bias relation down to scales of about $\gsim1$2 Mpc/h.
\end{abstract}

\begin{keywords}
(cosmology:) large-scale structure of Universe -- galaxies: clusters: general --
 catalogues -- galaxies: statistics
\end{keywords}

\section{Introduction}

The luminous matter we observe on the sky represents only a small fraction of the total matter in the Universe and yet with a careful treatment of the observational selection effects and the processes of galaxy formation we can hope to extract valuable information about the distribution of all matter from the distribution of luminous matter alone.  The more precise the techniques for making this connection are the better we will be able to test our theories for the history of the Universe. 

In 1934 Hubble found that the distribution of galaxy  counts in cells on the sky is well fitted by a lognormal distribution \citep{1934ApJ....79....8H}.  More recently, \citet{2005MNRAS.356..247W} showed that this model is valid at least down to gridding scales  of about 10 Mpc for galaxies in the 2DF catalogue.
As galaxies are good tracers of matter on large cosmological scales the lognormal model should also apply to the matter field at least to some degree. 
\citet[][]{kitaura_sdss}  showed recently that the matter field reconstruction using (least squares) Wiener filtering is very well fit by a lognormal distribution after smoothing the reconstruction with  a Gaussian kernel of radius $r_{\rm S}$  for  $10\,{\rm Mpc}\lsim1{r_{\rm S}}\lsim130\,{\rm Mpc}$.  

From a physical point of view, one would expect  the density field to be lognormally distributed after it has been smoothed on an appropriate scale.  This follows from assuming an initially Gaussian density and velocity field and extrapolating the continuity equation for the matter flow into the
nonlinear regime with the linear velocity fluctuations
\citep[see][]{1991MNRAS.248....1C}.  Since the lognormal field is not
able to describe caustics, we expect this distribution to fail below some
threshold smoothing scale. \cite{2001ApJ...561...22K} demonstrated
that the lognormal distribution is a good approximation up to
overdensties of about $\delta\sim$100.

Shortly after the success of Wiener filtering in large--scale structure reconstruction \citep[see][]{1995ApJ...449..446Z}, which assumes a Gaussian prior for the matter field, a reconstruction filter based on the lognormal prior distribution was proposed \citep[see][]{1995MNRAS.277..933S}. With such a filter nonlinearities in the density field should be better recovered.  In \citet[][]{1995MNRAS.277..933S} the Wiener filter was generalized to be applied to a lognormal distribution by a variable transformation. A problem with this approach is that the noise covariance has a complex form even for the simple Poisson likelihood assumption and is difficult to efficiently apply to realistic data--sets.

The idea of modeling the galaxies as Poisson--sampled from a lognormal underlying field was first applied to data by \citet{2000ASPC..218..181S}. They proposed to use a filtering scheme based on an expansion of the logarithm of the matter field as a sum of harmonics. The density reconstruction using this technique as presented by \citet{2000ASPC..218..141S} is fairly smooth and nonlinear structures cannot be easily recognised. This could be due to the sparse sampling of the PSCz catalogue which was used in their study or to the truncation of the harmonic series. 

As demonstrated in \citet[][]{kitaura}, the Poissonian likelihood can be easily regularised by combining it with a prior when estimating the maximum a posteriori. They showed this calculation for Gaussian and entropic prior distribution functions. 

The idea  of using the full Poissonian likelihood without remaining at second order approximations using only the noise covariance matrix  is  based on the Richardson--Lucy deconvolution algorithm  
\citep[see][]{1972JOSA...62...55R,1974AJ.....79..745L}.
\citet[][]{sheppvardi} showed that this filter comes from the maximum likelihood estimate of the Poissonian likelihood.
\citet[][]{1999MNRAS.303..179N} proposed using this method to recover the density field from the Lyman alpha forest. The problem that arose here was that the algorithm requires truncation as it assumes a flat prior for the matter field and thus the deconvolution of the response operator is not regularised.
However, as \citet[][]{kitaura} pointed out, this kind of problem can be solved by introducing a prior. \citet{ensslin} proposed to calculate higher order corrections to obtain an estimate for the mean of the posterior distribution by employing a generating functional formalism with a Poissonian process on top of a lognormal field for the galaxy distribution.

In this work we present a general expression for the Poisson--lognormal filter given an arbitrary nonlinear galaxy bias. We derive this filter as the maximum a posteriori solution  assuming a lognormal prior distribution for the matter field with a constant mean field and modeling the observed galaxy distribution by a Poissonian process. We have performed a three--dimensional implementation of this filter  with a very efficient Newton--Krylov inversion scheme extending the \textsc{argo} computer code to perform nonlinear inversions \citep[see][]{kitaura}. Furthermore, we have tested it for a linear galaxy bias relation and compared the results  with other  density field estimators commonly used in the literature (e.g.~the inverse weighting scheme and the least squares (LSQ) Wiener filter). 
The one--dimensional lognormal probability distribution is known to fit the matter distribution well up to overdensities of about $\delta\sim$100 as found by \citet{2001ApJ...561...22K}. Our results show, however, good agreement for overdensities even above $\delta\sim$1000 which exceeds by one order of magnitude the expected regime in which the lognormal is expected to be valid. 
The reason for this apparent disagreement is that for the filter presented here the lognormal assumption enters as a prior distribution function, but the maximum a posteriori solution is also conditioned on the data. For the same reason \citet[][]{kitaura_sdss} obtained a highly non--Gaussian distributed matter field after using LSQ--Wiener filtering which according to the Bayesian formalism assumes a Gaussian distribution. We find that the Poisson--lognormal filter has a range of  applicability in recovering matter density fields down to  scales of about $\gsim1$2 Mpc/h. However, the matter statistics show that the Poisson--lognormal filter fails to recover underdense regions $\delta\lsim1-0.6$ with very few data.
In addition, we test the maximum a posteriori assuming a Gaussian prior and found that it is not capable of recovering the density field when $\delta\gg1$ and gives negative densities in low density regions which makes this filter unreliable for recovering densities of $\delta\lsim11$.

Finally, we show in appendix \ref{app:LSQ} that the LSQ filter is the optimal linear filter under a Poisson noise assumption using up to second order statistics and does not neglect any signal to noise correlation, contrary to what has been assumed in the literature \citep[see for example][]{1995ApJ...449..446Z,1998ApJ...503..492S,2004MNRAS.352..939E,kitaura}.
We also derive in appendix \ref{app:LPGL} a filter with a lognormal model for the underlying signal and an additive, signal--independent and  Gaussian distributed noise which could be of interested in other fields of astronomy.

The paper is structured as follows.  In section \ref{sec:bayes} we present the Bayesian approach used in this work. After defining the likelihood for the galaxy sample and the prior distributions for the matter field  we calculate the maximum a posteriori (MAP) estimates for the underlying density field. Then the numerical scheme is presented in section \ref{sec:numerics} which permits us to solve the MAP estimates. We then present in section \ref{sec:results} a series of numerical experiments which show the performance of the different density estimators. Finally, we discuss our results.

\begin{figure}
\begin{tabular}{c}
\includegraphics[width=7.5cm]{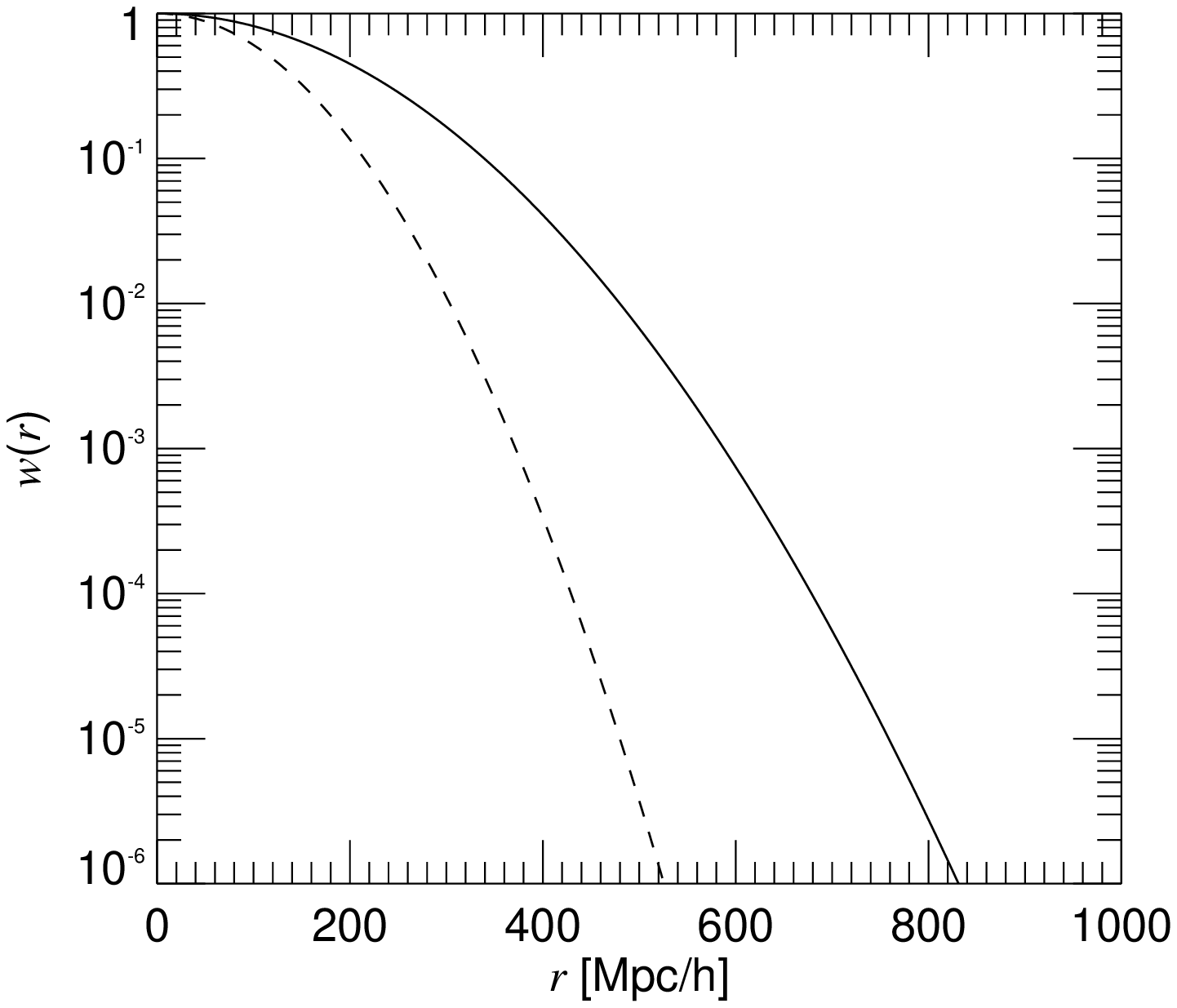}
\put(-220,170){\rotatebox[]{0}{ $\times10^{-4}$}}
\put(-100,130){\rotatebox[]{0}{ $w_1$}}
\put(-115,100){\rotatebox[]{0}{ $w_2$}}
\end{tabular}
\caption{Two models of completeness $w$ emulating apparent magnitude limit effects (continuous curve: $w_1$ and dashed curve: $w_2$) dependent on the distance $r$ to the observer in Mpc/h.} 
\label{fig:sel}
\end{figure}

\section{Bayesian approach}
\label{sec:bayes}

A Bayesian approach requires the definition of a likelihood and a prior. A full Bayesian analysis would require the complete characterization of the posterior distribution using sampling schemes \citep[see e.g.~][]{2004PhRvD..70h3511W}. We leave such an approach for a forthcoming publication and restrict ourselves here to calculate the extrema which leads to the maximum a posteriori expressions. This permits us to get a fast estimate of the density field.
In this work, we  consider a Poissonian likelihood for the observed distribution of galaxies and combine it with a Gaussian and a lognormal prior distribution for the overdensity field. 
In the next subsections these distribution functions are presented and the calculation of the different MAP--estimators are shown in detail.

\subsection{Poissonian likelihood}

The likelihood represents the observation process which leads to the
data. It is the probability distribution function that describes the
nature of the observable. In this case we look for a model that
accounts for the discrete nature of a galaxy distribution, the
so--called shot noise. This kind of noise is traditionally modeled by
a Poissonian distribution \citep[see for example][]{Peebles-80}.  Such
a model assumes that each cell of the Universe in which we count some
number of galaxies (maybe according to a certain luminosity type) is
statistically independent from each other. However, the variance of
counts in cells including a correlation term predicts the
non--Poissonian character of the distribution of
galaxies \citep[see][]{Peebles-80}.  Hierarchical structure formation
models assume that galaxies form inside dark matter halos via the
energy dissipation by baryons \citep[see
e.~g.~][]{1978MNRAS.183..341W}.  \citet[][]{2001MNRAS.320..289S}
showed based on numerical N--body simulations that in regions of lower
than average overdensity, the scatter in the halo biasing (the
relation between the dark matter halos and the underlying dark matter
distribution) is generally smaller than the mean Poisson shot noise,
and in overdense regions it is larger \citep[for a review on the halo
model see][]{cooray-2002-372}.  \citet[][]{1996MNRAS.282..347M}
already pointed out that halo--exclusion can cause sub--Poisson
variance.
\citet[][]{2002MNRAS.333..730C} demonstrated with higher resolved N--body simulations that the galaxy biasing
process, as well as the halo biasing process, is not only determined
by the local value of the mass density field, but also by other local
quantities, such as clumpiness, and by non--local properties, such as the
large--scale tidal field.
Accounting for all these effects is out of scope of this work, but should certainly be further investigated.

Here, we will restrict ourselves to a model in which the
observed distribution of galaxies is given by an inhomogeneous Poisson
realization of a continuous density field.  We define the likelihood function
as \citep[see][]{1980lssu.book.....P}:
\begin{equation}
 {\cal L}(\langle\mbi N^{\rm o}_{{\rm g}}\rangle_{\rm g}|\mbi N^{\rm o}_{{\rm g}})=\Pi_{i=1}^{N_{\rm cells}}\exp\left[-\langle N^{\rm o}_{{\rm g},i}\rangle_{\rm g}\right]\frac{\langle N^{\rm o}_{{\rm g},i}\rangle_{\rm g}^{N^{\rm o}_{{\rm g},i}}}{N^{\rm o}_{{\rm g},i}!}{,}
\end{equation} 
with  $N^{\rm o}_{{\rm g},i}$ denoting the number count of observed galaxies in cell $i$, and $N_{\rm cells}$ being the total number of cells.
 Here $\langle \{ \, \} \rangle_{\rm g} \equiv \langle \{ \, \} \rangle_{(N^{\rm o}_{\rm g}\mid \lambda^{\rm o})} \equiv \sum^{\infty}_{N^{\rm o}_{\rm g}=0} \, P_{\rm Pois}(N^{\rm o}_{\rm g}\mid w\lambda) \{ \, \}$ denotes an ensemble average over the
Poissonian distribution with the expected number of galaxy counts given by the Poissonian
ensemble average: $\lambda^{\rm o}\equiv w\lambda\equiv\langle N^{\rm o}_{\rm g}\rangle_{\rm g}$. The expected number count is related to the underlying continuous galaxy overdensity field $\delta_{{\rm g},i}$ through:
\begin{equation}
\langle N^{\rm o}_{{\rm g},i}\rangle_{\rm g}\equiv\overline{N}_{\rm g}w_{i}(1+\delta_{{\rm g},i}){,}
\end{equation} 
where $\overline{N}_{\rm g}$ is the mean number count of galaxies and $w_{i}$ the completeness at cell $i$.
The logarithm of the likelihood can be written as:
\begin{equation}
\ln {\cal L}_i= -\overline{N}_{\rm g}w_{i}(1+\delta_{{\rm g},i})+N^{\rm o}_{{\rm g},i}\ln(\overline{N}_{\rm g}w_{i}(1+\delta_{{\rm g},i}))-\ln(N^{\rm o}_{{\rm g},i}!){.}
\end{equation}

\subsection{Gaussian prior}

The prior probability distribution function describes the statistical nature of the signal one wants to infer from the observed data. Here the physical model of the underlying matter field comes in. As inflationary scenarios predict a close to Gaussian distribution function for the initial density fluctuations \citep[see][]{1981PhRvD..23..347G, 1982PhRvL..49.1110G, 1982PhLB..117..175S, 1982CMaPh..87..395H,  1982PhLB..108..389L, 1982PhRvL..48.1220A, 1983PhRvD..28..679B} and linear theory preserves this property throughout cosmic evolution it is reasonable to assume a Gaussian prior to model the large--scale matter field. Note, however, that this can only be true for $|\delta|\ll1$ since otherwise the Gaussian distribution predicts unphysical negative densities. 
Here we follow \citet[][]{1986ApJ...304...15B} to describe the prior probability distribution of the density field by a multivariate Gaussian distribution function:
\begin{equation}
{\cal P}(\mbi \delta_{\rm M}|\mbi p)= \frac{1}{\sqrt{(2\pi)^{N_{\rm cells}}\det(\mat S)}}\exp\left[-\frac{1}{2}\mbi \delta_{\rm M}^\dagger\mat S^{-1}\mbi \delta_{\rm M}\right]{,}
\end{equation} 
with $\mbi p$ being the set of cosmological parameters which determine the autocorrelation matrix $\mat S$ and $\delta_{\rm M}$ is the overdensity in mass.
The application of the autocorrelation matrix  $\mbi S$ to a vector $\mbi x$ is a convolution of the form:  $\mat S\mbi x\equiv S(r) \circ x (r)$ (with $r$ being the cell coordinates in configuration space of the box  $r=\{i_X,i_Y,i_Z\}$ and with "$\circ$" denoting the convolution operation).

Note that the Fourier transform of the autocorrelation matrix $\mat S$ is equal to the power spectrum:
$\hat{\hat{\mat S}}(\mbi k,\mbi k')\equiv (2\pi)^3P(\mbi k')\delta_{\rm D}(\mbi
k-\mbi k')$ \citep[using the same Fourier definitions as in][]{kitaura}.
The logarithm of the prior distribution function can be written as:
\begin{equation}
\ln {\cal P}(\mbi \delta_{\rm M}|\mbi p)= -\frac{1}{2}\mbi \delta_{\rm M}^\dagger\mat S^{-1}\mbi \delta_{\rm M}+c{,}
\end{equation} 
with $c$ being the logarithm of the normalization.

The posterior distribution function $P$ is proportional to the product of the prior $\mathcal P$ and the likelihood $\mathcal L$. To find the maximum a posteriori (MAP) we need to calculate the extremum. Performing the derivative of the posterior with respect to the matter overdensity field $\delta_{\rm M}$ yields:
\begin{equation}
\frac{\partial \ln P}{\partial \mbi \delta_{\rm M}}\propto\frac{\partial \ln {\cal P}}{\partial \mbi \delta_{\rm M}}+\frac{\partial\ln {\cal L}}{\partial \mbi \delta_{\rm M}}=0{.}
\end{equation} 
The derivative of the prior leads to:
\begin{equation}
\frac{\partial \ln {\cal P}}{\partial \mbi \delta_{\rm M} }=-\mat S^{-1}\mbi \delta_{\rm M}{.}
\label{eq:priorder}
\end{equation} 
Since the likelihood is expressed as a function of the galaxy density field, we need to define the bias between the galaxy and matter fields.

\subsubsection{Linear bias}

As a particular case, let us consider a linear  bias function given by:
\begin{equation}
\delta_{{\rm g},i}=\sum_jb_{i,j}\delta_{{\rm M},j}{,}
\end{equation} 
 which relates the corresponding power spectra in the following way: $\hat{b}(\mbi k)=\sqrt{P_{\rm g}(\mbi k)/P_{\rm M}(\mbi k)}$, with $P_{\rm g}(\mbi k)$ being the galaxy power--spectrum and $P_{\rm M}(\mbi k)$ being the matter power--spectrum. 

The derivative of the likelihood with respect to the matter overdensity fields is then given by:
\begin{equation}
\sum_i\frac{\partial\ln {\cal L}_i}{\partial \delta_{{\rm M},k}}=\sum_ib_{i,k}\left[-\overline{N}_{\rm g}w_{i} +\frac{N^{\rm o}_{{\rm g},i}}{1+\sum_l b_{i,l}\delta_{{\rm M},l}}\right]{.}
\end{equation} 
Adding this result to the prior term Eq.~(\ref{eq:priorder}) we obtain the MAP equation:
\begin{equation}
\delta_{{\rm M},i}^{\rm G}=\sum_jS_{i,j}\sum_lb_{l,j}\left(\frac{N^{\rm o}_{{\rm g},l}}{1+\sum_k b_{l,k}\delta_{{\rm M},k}^{\rm G}}-\overline{N}_{\rm g}w_{l}\right){,}
\label{eq:MAPG1}
\end{equation} 
with the superscript ${\rm G}$ standing for the Gaussian prior assumption.

\subsubsection{Unity bias}

Let us consider the special case when the matter field is equal to a {\it continuous} galaxy field:
\begin{equation}
\delta_{{\rm g},i}=\delta_{{\rm M},i}{,}
\end{equation} 
then the MAP equation reads:
\begin{equation}
\delta_{{\rm M},i}^{\rm G}=\sum_jS_{i,j}\left(\frac{N^{\rm o}_{{\rm g},j}}{1+\delta_{{\rm M},j}^{\rm G}}-\overline{N}_{\rm g}w_{j}\right){.}
\label{eq:MAPG}
\end{equation}

\begin{figure*}
\begin{tabular}{cc}
\includegraphics[width=7.cm]{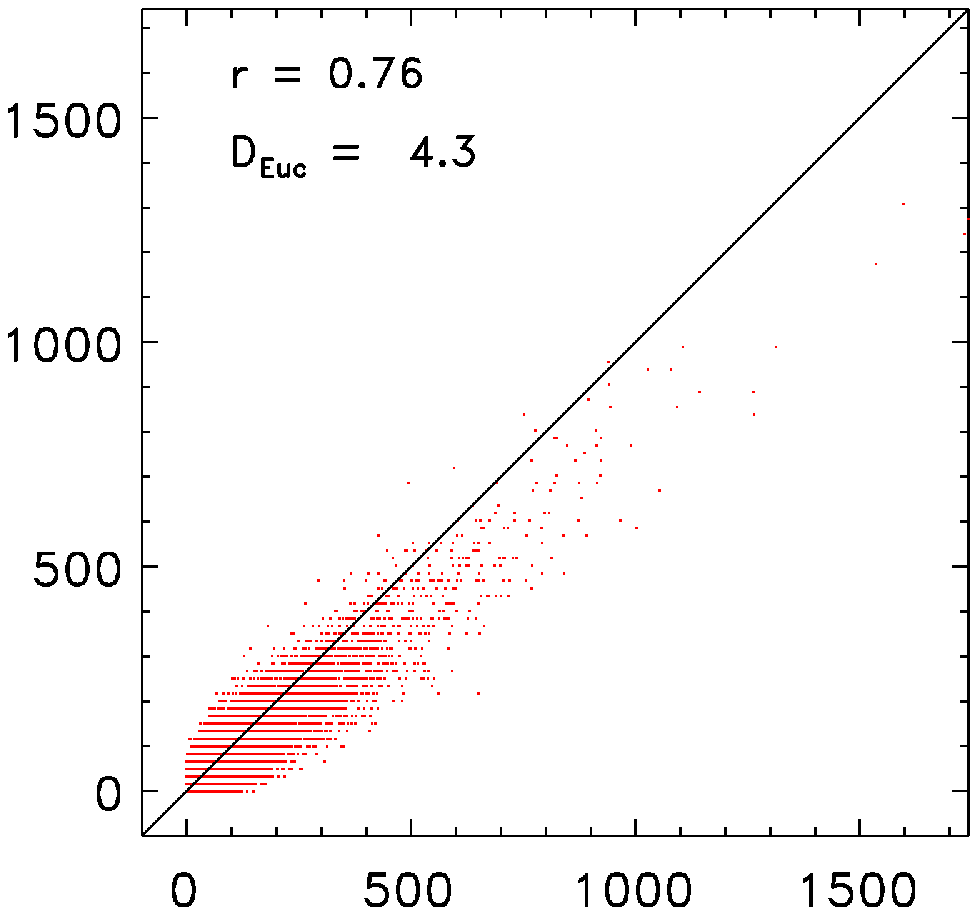}
\put(-50,160){\rotatebox[]{0}{\large$256^3$}}
\put(-90,-10){\rotatebox[]{0}{\large$\delta_{\rm M}$}}
\put(-210,100){\rotatebox[]{90}{\large$\delta_{\rm g}$}}
\hspace{1.5cm}
\includegraphics[width=7.cm]{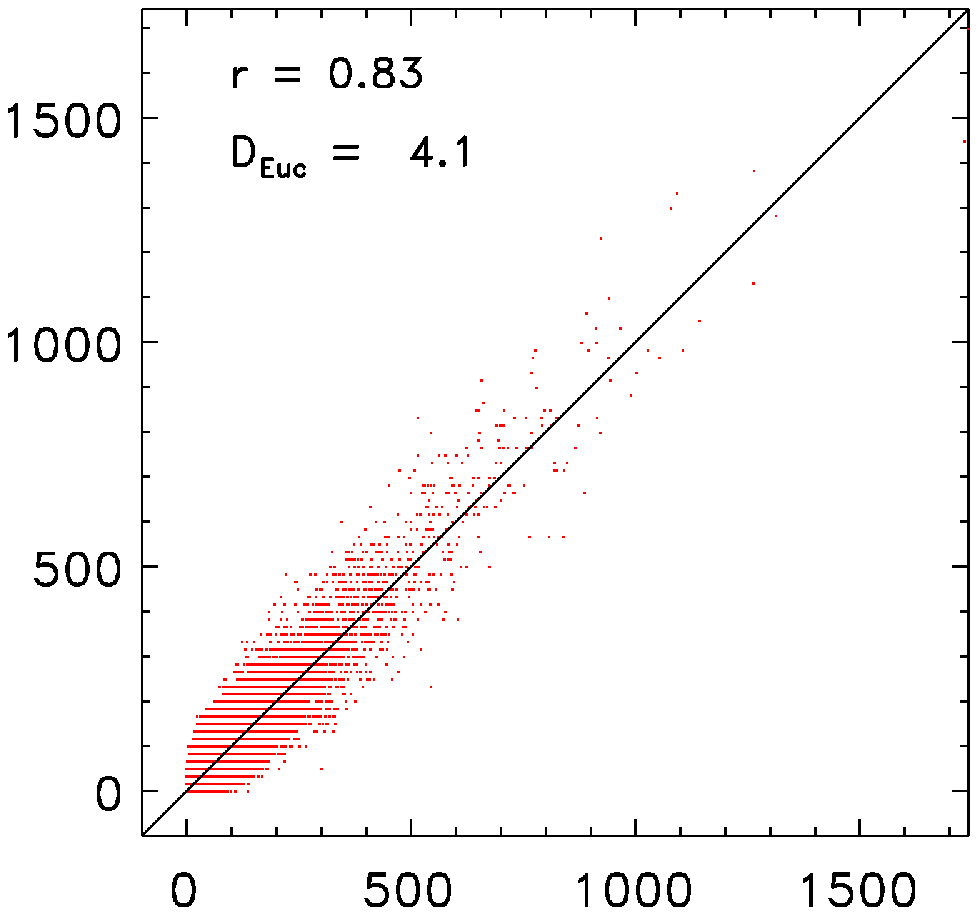}
\put(-50,160){\rotatebox[]{0}{\large$256^3$}}
\put(-90,-10){\rotatebox[]{0}{\large$\delta_{\rm M}$}}
\put(-210,100){\rotatebox[]{90}{\large$\delta_{\rm g}$}}\\
\end{tabular}
\caption{Cell-to-cell overdensity correlation between  the mock galaxy
  sample and the matter field based on the Millenium run
  \citep[][]{2005Natur.435..629S}. On the left: mock galaxy catalogue
  by \citet[][]{DeLucia-Blaizot-07}. On the right: Poisson sample over
  the matter field with a constant completeness of $10^{-4}$. Also
  given are the correlation coefficient $r$ and Euclidean distance to
  the underlying matter field D$_{\rm Euc}$.} 
\label{fig:corr_gal}
\end{figure*}

\subsection{Lognormal prior}

Here we introduce the lognormal prior distribution as proposed by \citet[][]{1991MNRAS.248....1C}:
\begin{equation}
{\cal P}(\mbi s|\mbi p)= \frac{1}{\sqrt{(2\pi)^{N_{\rm cells}}\det(\mat S_{\rm L})}}\exp\left[-\frac{1}{2}(\mbi s-\mbi \mu)^\dagger\mat S^{-1}_{\rm L}(\mbi s-\mbi \mu)\right]{,}
\end{equation} 
with $\mbi s$ being the logarithm of the weighted matter density:
\begin{equation}
s_i\equiv\log(1+\delta_{{\rm M}i}){,}
\end{equation} 
and  $\mat S_{\rm L}$ the corresponding autocorrelation matrix.
Note that the lognormal autocorrelation function $\mat S_{{\rm L}}$ applied to a vector $\mbi x$ is again a convolution: $\mat S_{{\rm L}}\mbi x\equiv S_{{\rm L}}(r)\circ x(r)$.
The transformation of the correlation function corresponding to the overdensity field to the signal $s$ is given by:
\begin{equation}
S_{{\rm L}}(r)\equiv\log(1+S(r)){.}
\end{equation} 
The mean field $\mbi \mu$ is taken to be:
\begin{equation}
\label{eq:mu}
\mu_i \equiv -\sigma_0^2/2{,}
\end{equation} 
with $\sigma_0^2\equiv S_{\rm L}(0)$ as used by \citet[][]{2001ApJ...561...22K} to ensure an overdensity field with zero mean\footnote{Here generalized to a multivariate lognormal distribution.}  \citep[for a formal derivation see][]{1991MNRAS.248....1C}.
The logarithm of the prior distribution yields:
\begin{equation}
\ln {\cal P}(\mbi s|\mbi p)= -\frac{1}{2}(\mbi s-\mbi \mu)^\dagger\mat S^{-1}_{\rm L}(\mbi s-\mbi \mu)+c{,}
\end{equation} 
with $c$ being some constant term.
In this case, we look for the extremum with respect to the signal $\mbi s$:
\begin{equation}
\frac{\partial \ln P}{\partial \mbi s}=\frac{\partial \ln {\cal P}}{\partial \mbi s}+\frac{\partial\ln {\cal L}}{\partial \mbi s}=0{.}
\end{equation} 
The derivative of the prior  has now an additional term due to the mean--field $\mbi \mu$:
\begin{equation}
\frac{\partial \ln {\cal P}}{\partial \mbi s}=-\mat S^{-1}_{\rm L}\left(\mbi s-\mbi \mu\right){.}
\label{eq:priorlog}
\end{equation} 
Now we need to relate the galaxy field to the matter field in order to express the likelihood as a function of the signal statistically defined through  the prior distribution function.

\subsection{General nonlinear bias}

Let us consider here a general nonlinear relation between the galaxy field and the  matter field. 
\begin{equation}
\delta_{{\rm g},i}=B(\mbi\delta_{{\rm M}})_i{,}
\end{equation} 
The derivative of the likelihood with respect to the signal $\mbi s$ which we want to recover can be written as:
\begin{equation}
\sum_i\frac{\partial\ln {\cal L}_i}{\partial s_k}=\sum_i\sum_l\frac{\partial\ln {\cal L}_i}{\partial \ln (1+\delta_{{\rm g},l})}\frac{\partial \ln (1+\delta_{{\rm g},l})}{\partial \ln (1+\delta_{{\rm M},{k}})}{.}
\label{eq:der}
\end{equation} 
The derivative of the likelihood with respect to the logarithm of the normalized galaxy field $\ln(1+\delta_{{\rm g},l})$ yields:
\begin{equation}
\frac{\partial\ln {\cal L}_i}{\partial \ln (1+\delta_{{\rm g},l})}=\left(N^{\rm o}_{{\rm g},i}-\overline{N}_{\rm g}w_{i}\left(1+B(\mbi\delta_{{\rm M}})_i\right)\right)\delta^{\rm K}_{i,l}{.}
\end{equation} 
The factor relating the galaxy field to the matter field yields:
\begin{eqnarray}
\frac{\partial \ln (1+{\delta}_{{\rm g},l})}{\partial \ln (1+{\delta}_{{\rm M},k})}&=&\frac{\partial \ln (1+B\left(\exp(\mbi s)-\vec{1}\right)_l)}{\partial s_{k}}\nonumber\\
&=&\frac{\partial B(\mbi\delta_{{\rm M}})_l}{\partial \delta_{{\rm M},{k}}}\frac{(1+\delta_{{\rm M},k})}{1+B(\mbi\delta_{{\rm M}})_l}{.}
\end{eqnarray} 
The final result for the derivative of the likelihood with respect to the logarithm of the normalized matter field $\mbi s$ assuming a general nonlinear bias is given by:
\begin{eqnarray}
\lefteqn{\sum_i\frac{\partial\ln {\cal L}_i}{\partial s_k}=}\\
&&\hspace{-1cm}\sum_i\frac{\partial B\left(\mbi\delta_{{\rm M}}\right)_i}{\partial \delta_{{\rm M},{k}}}\frac{(1+\delta_{{\rm M},k})}{1+B(\mbi\delta_{{\rm M}})_i}\left(-\overline{N}_{\rm g}w_{i} \left(1+B(\mbi\delta_{{\rm M}})_i\right)+N^{\rm o}_{{\rm g},i}\right)\nonumber{.}
\end{eqnarray} 
Combining this result with the prior term (Eq.~\ref{eq:priorlog}) we obtain the MAP equation:
\begin{eqnarray}
\lefteqn{\sum_j S^{-1}_{{\rm L}{i,j}}\left(\ln(1+\delta_{{\rm M},j}^{\rm L})-\mu_j\right)=}\\
\label{eq:MAPLN}
&&\hspace{0cm}\sum_l\frac{\partial B\left(\mbi\delta_{{\rm M}}^{\rm L}\right)_l}{\partial \delta_{{\rm M},{i}}^{\rm L}}\frac{(1+\delta_{{\rm M},i}^{\rm L})}{1+B(\mbi\delta_{{\rm M}}^{\rm L})_l}\times\left.\left(N^{\rm o}_{{\rm g},l}-\overline{N}_{\rm g}w_{l}\left(1+B(\mbi\delta_{{\rm M}}^{\rm L})_l\right)\right)\right)\nonumber{,}
\end{eqnarray} 
with the superscript $\rm L$ standing for the lognormal prior.

\subsubsection{Linear bias}
\label{sec:linbias}

For the linear bias case the derivative of the likelihood reduces to the following expression:
\begin{eqnarray}
\sum_i\frac{\partial\ln {\cal L}_i}{\partial s_k}&=&\sum_i \frac{b_{i,k}(1+\delta_{{\rm M},k})}{1+\sum_{l'}b_{i,l'}\delta_{{\rm M},{l'}}}\\
&&\times\left(-\overline{N}_{\rm g}w_{i} \left(1+\sum_jb_{i,j}\delta_{{\rm M},{j}}\right)+N^{\rm o}_{{\rm g},i}\right)\nonumber{.}
\end{eqnarray} 
Accordingly, the MAP equation reads:
\begin{eqnarray}
\lefteqn{\sum_j S^{-1}_{{\rm L}{i,j}}\left(\ln\left(1+\delta_{{\rm M},j}^{\rm L}\right)-\mu_j\right)=}\\
&&\sum_k \frac{b_{k,i}(1+\delta_{{\rm M},k}^{\rm L})}{1+\sum_{l'}b_{k,l'}\delta_{{\rm M},{l'}}^{\rm L}}\left(N^{\rm o}_{{\rm g},k}-\overline{N}_{\rm g}w_{k}\left(1+\sum_lb_{k,l}\delta_{{\rm M},{l}}^{\rm L}\right)\right)\nonumber{.}
\end{eqnarray}

\subsubsection{Unity bias}

For an unity bias the derivative of the likelihood reduces to:
\begin{equation}
\delta_{{\rm g},i}=\delta_{{\rm M},i}{.}
\end{equation} 
Then, the derivative of the likelihood reduces to:
\begin{equation}
\sum_i\frac{\partial\ln {\cal L}_i}{\partial s_k}=-\overline{N}_{\rm g}w_{k} \exp\left(s_{k}\right)+N^{\rm o}_{{\rm g},k}{,}
\end{equation} 
and the MAP equation reads:
\begin{equation}
\sum_jS^{-1}_{{\rm L}{i,j}}\left(s_j-\mu_j\right)=N^{\rm o}_{{\rm g},i}-\overline{N}_{\rm g}w_{i}\exp(s_i){.}
\end{equation} 
Using the definitions: $\overline{N}_{\rm g}w_{j}\exp(s_j)=\overline{N}_{\rm g}w_{j}(1+\delta_{{\rm g},i})=\langle N^{\rm o}_{{\rm g}}\rangle_{\rm g}$ and $\epsilon^{\rm o}_j\equiv N^{\rm o}-\langle N^{\rm o}_{{\rm g}}\rangle_{\rm g}$ we can rewrite the MAP equation as:
\begin{equation}
\sum_jS^{-1}_{{\rm L}{i,j}}\left(s_j-\mu_j\right)=\epsilon^{\rm o}_i{,}
\end{equation} 
The signal $\mbi s$ is thus given by the propagation of the noise $\mbi \epsilon^{\rm o}$ given the correlation $\mat S_{{\rm L}}$ up to a shift due to the mean $\mbi \mu$.
Expressing it as a function of the matter overdensity-field we obtain:
\begin{equation}
\label{eq:MAPLNUB}
\sum_jS^{-1}_{{\rm L}{i,j}}\left(\ln\left(1+\delta_{{\rm M},j}^{\rm L}\right)-\mu_j\right)=\left(N^{\rm o}_{{\rm g},i}-\overline{N}_{\rm g}w_{i}(1+\delta_{{\rm M},i}^{\rm L})\right){.}
\end{equation}

\begin{table*}
\rotatebox[]{0}{
\begin{tabular}{|c|c|c|} 
& \hspace{6cm} {\bf LIKELIHOODS}&\\
    &     {\bf Gaussian} & {\bf Poissonian}     \\ \hline
{\bf PRIORS}  &  &\\ 
 {\bf Flat}  & (a)  &  $\delta_{{\rm g},i}^{\rm IW}=\frac{ N_{{\rm g},i}^{\rm o}}{w_i\overline{N_{\rm g}}}-1$\\
{\bf Gaussian} & ${\delta}_{{\rm M},i}^{\rm LSQ}=\sum_j \left(S^{-1}_{i,j}+w_j\overline{N}_{\rm g}\delta^{\rm K}_{i,j}\right)^{-1}\overline{N}_{\rm g}\delta_{{\rm g},j}^{{\rm o}}$&$\delta_{{\rm M},i}^{\rm G}=\sum_jS_{i,j}\left(\frac{N^{\rm o}_{{\rm g},j}}{1+\delta_{{\rm M},j}^{\rm G}}-\overline{N}_{\rm g}w_{j}\right)$\\
 {\bf Lognormal} &  (b)    & $\sum_jS^{-1}_{{\rm L}{i,j}}\left(\ln\left(1+\delta_{{\rm M},j}^{\rm L}\right)-\mu_j\right)=N^{\rm o}_{{\rm g},i}-\overline{N}_{\rm g}w_{i}(1+\delta_{{\rm M},i}^{\rm L})$        \\ 
\end{tabular}
}
\caption{ \label{tab:filters} Filters which are used in this work classified by the assumed likelihood and prior (with the exception of (a) and (b)). Note, that the bias has been set to one. (a) COBE-filter used in CMB mapping \citep[see][]{1992issa.proc..391J}. (b) for a derivation of this filter see appendix \ref{app:LPGL}.
}
\end{table*}

\section{Numerical approach}
\label{sec:numerics}

The problem we are studying here requires the solution of a nonlinear system of $256^3$ (about $17\cdot10^6$) coupled equations. Note, that each cell introduces an equation.
Thus, to find the MAP solution (Eqns.~\ref{eq:MAPG} and \ref{eq:MAPLN}) we apply an operator based iterative inversion scheme as proposed in \citet[][]{kitaura} which reduces the most expensive operations to  FFTs.
In particular,  we use a nonlinear Newton--Krylov scheme which is briefly presented in the next subsections \citep[for a reference see e.g.~][]{kitaura}. 

\subsection{Method}

Let us write a system of nonlinear equations as:
$A(\mbi x)=\mbi f$, with $\mbi A$ being the nonlinear operator dependent on $\mbi x$ and $\mbi f$ some constant vector.
We then define the gradient of the quadratic approximation as:
\begin{equation}
\nabla  Q(\mbi x)\equiv A(\mbi x)-\mbi f{.}
\end{equation} 
The corresponding Hessian matrix is then given by the second derivative of the gradient of $Q$:
\begin{equation}
\mat H\equiv \nabla\nabla  Q(\mbi x){.}
\end{equation} 
The basic Newton--Raphson solver scheme is given by:
\begin{equation}
\mbi x^{j+1}=\mbi x^{j}-\left(\mat H^{j}\right)^{-1}\nabla  Q(\mbi x^{j}){.}
\end{equation} 
This scheme turns out to be extremely inefficient. Therefore, we implement a Krylov step in which the solution is updated in the following way:
\begin{equation}
\mbi x^{j+1}=\mbi x^{j}+\tau^j\xi^j{,}
\end{equation} 
with the stepsize $\tau$ given by \citep[for a derivation see][]{kitaura}:
\begin{equation}
\tau^j=-\frac{\hspace{0.cm}\mbi \xi^{j\dagger}\nabla Q(\mbi x^j)}{\hspace{0.cm}\mbi\xi^{j\dagger}{\mat H^j}\mbi\xi^j}{.}
\label{eq:New5}
\end{equation}
and $\xi$ being the searching vector \citep[for schemes to calculate $\xi$ see][]{kitaura}.
In the next subsections we give the particular expressions for the quantities required to calculate the MAP given a Gaussian prior first and finally given a lognormal prior.

\subsubsection{Gaussian prior}

Rewriting Eqn.~\ref{eq:MAPG} as:
\begin{equation}
\mbi \delta_{{\rm M}}^{\rm G}-\mat S\,{\rm diag}({\mbi 1+\mbi \delta_{{\rm M}}^{\rm G}})^{-1}{\mbi N^{\rm o}_{{\rm g}}}=\overline{N}_{\rm g}\mat S\mbi w_{{\rm g}}{.}
\end{equation} 
We can identify $A(\mbi s)=\mbi \delta_{{\rm M}}^{\rm G}-\mat S\,{\rm diag}({\mbi 1+\mbi \delta_{{\rm M}}^{\rm G}})^{-1}{\mbi N^{\rm o}_{{\rm g}}}$ and $\mbi f=\overline{N}_{\rm g}\mat S\mbi w_{{\rm g}}$.
The corresponding gradient of the quadratic form is given by:
\begin{equation}
\nabla  Q(\mbi s)=\mbi \delta_{{\rm M}}^{\rm G}-\mat S\,\left({\rm diag}({\mbi 1+\mbi\delta_{{\rm M}}^{\rm G}})^{-1}{\mbi N^{\rm o}_{{\rm g}}}-\overline{N}_{\rm g}\mbi w_{{\rm g}}\right){,}
\end{equation} 
and the Hessian yields:
\begin{equation}
\mat H={\1}+\mat S\,{\rm diag}({\mbi 1+\mbi\delta_{{\rm M}}^{\rm G}})^{-2}{\mbi N^{\rm o}_{{\rm g}}}{.}
\end{equation}

\subsubsection{Lognormal prior}

We formulate Eq.~\ref{eq:MAPLNUB} in an analogous way to the previous subsection as:
\begin{equation}
\mat S_{\rm L}^{-1}\left(\mbi s-\mbi\mu\right)+{\rm diag}\left(\overline{N}_{\rm g}\mbi w_{\rm g}\right)\exp(\mbi s)=\mbi N^{\rm o}_{\rm g}{,}
\end{equation} 
with $A(\mbi s)=\mat S_{\rm L}^{-1}\left(\mbi s-\mbi\mu\right)+{\rm diag}\left(\overline{N}_{\rm g}\mbi w_{\rm g}\right) \exp(\mbi s)$ and $\mbi f= \mbi N^{\rm o}_{\rm g}$.
Thus, the gradient of the quadratic form is given by:
\begin{equation}
\nabla  Q(\mbi s)=\mat S^{-1}_{\rm L}\left(\mbi s-\mbi\mu\right)+{\rm diag}\left(\overline{N}_{\rm g}\mbi w_{\rm g}\right)\exp(\mbi s)-\mbi N^{\rm o}_{\rm g}{,}
\end{equation} 
and the corresponding Hessian matrix reads:
\begin{equation}
\mat H=\mat S^{-1}_{\rm L}+{\rm diag}\left(\overline{N}_{\rm g} \mbi w_{\rm g}\right){\rm diag}\left(\exp(\mbi s)\right){.}
\end{equation}

\section{Numerical experiments}

\label{sec:results}

In this section we investigate the performance of the Poisson--lognormal filter.  
 We construct the mock  observed galaxy distributions by making a Poisson sample over the dark matter particles of the Millennium run according to different completeness models (see subsection \ref{sec:input}). This permits us to avoid the galaxy biasing and redshift distortions problems in our tests. 

We also test the Poisson--lognormal filter against other filters. We make a comparison to the MAP with a Gaussian prior assumption, to the inverse weighted data, and to the LSQ--Wiener filter (see section below \ref{sec:filters}). 
For an overview of the filters used in this work see table \ref{tab:filters}.

Finally, we test the quality of the reconstruction by making a cell--to--cell comparison  to the underlying  matter field which is assumed to be given by the dark matter distribution of  the Millennium run at redshift zero \citep[see][]{2005Natur.435..629S}. In addition, we study the matter density statistics of the dark matter field and the reconstructions.


\begin{figure*}
\begin{tabular}{cc}
\includegraphics[width=8.cm]{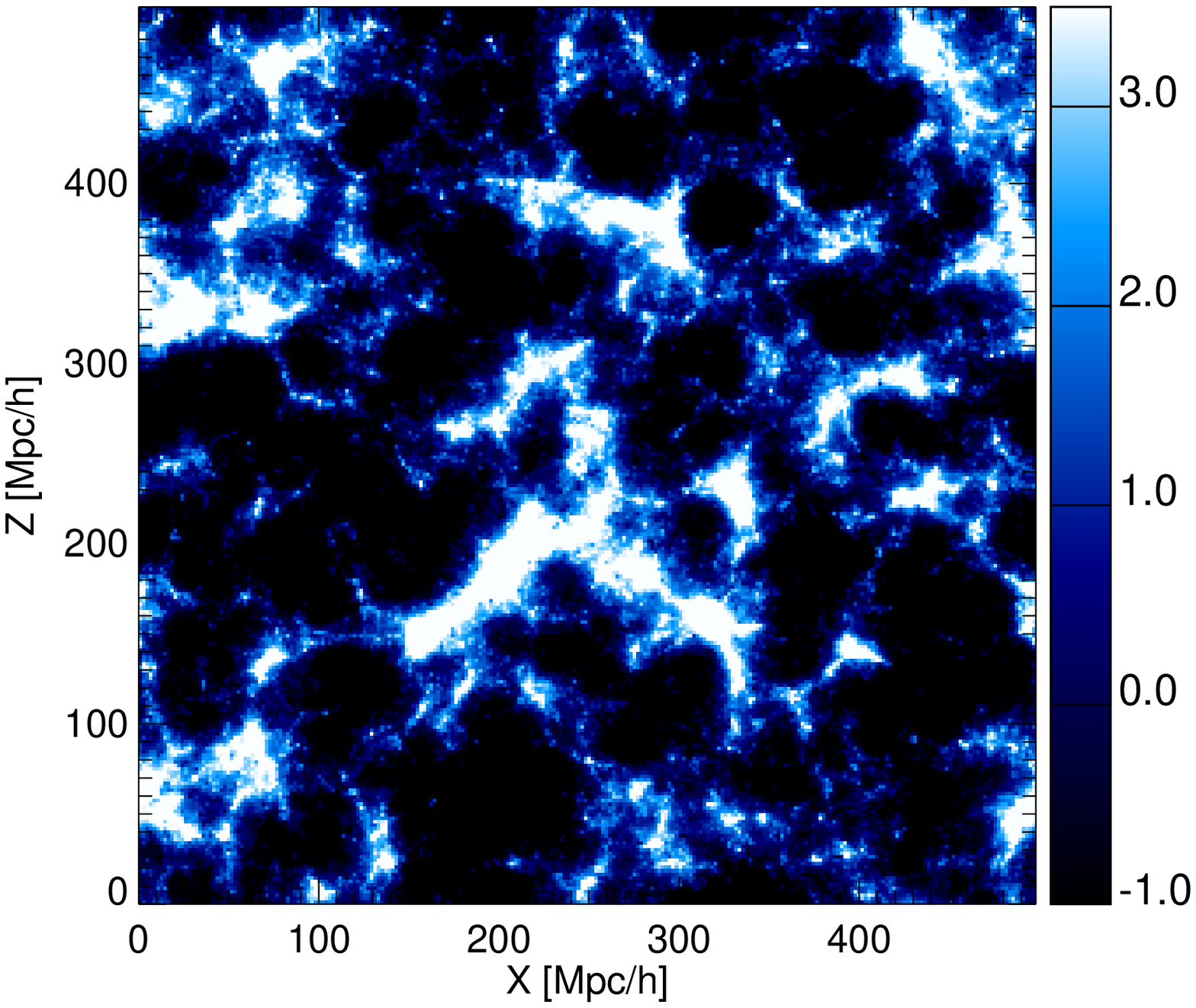}
\put(-5,100.){\rotatebox[]{0}{$\delta_{\rm M}^{\rm G}$}}
\hspace{1.5cm}
\includegraphics[width=6.cm]{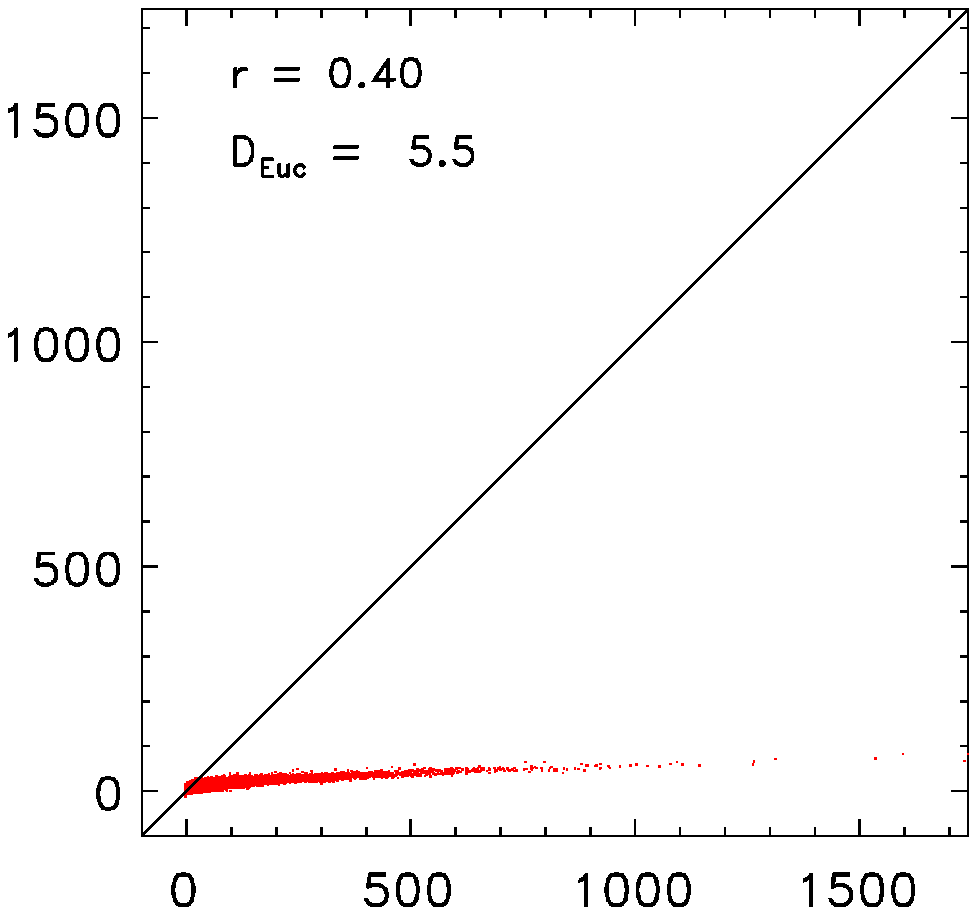}
\put(-50,160){\rotatebox[]{0}{\large$256^3$}}
\put(-185,100.){\rotatebox[]{90}{$\delta_{\rm M}^{\rm G}$}}
\put(-80,10){\rotatebox[]{0}{\large$\delta_{\rm M}$}}\\
\end{tabular}
\caption{Reconstruction assuming a Gaussian prior. Left panel: mean over 15 neighboring slices around Y$\sim176$ Mpc through a 500 Mpc cube box with a $256^3$ grid without smoothing. Right panel:  cell--to--cell correlation between the overdensity of the full three--dimensional reconstruction and the matter field.} 
\label{fig:gapmap}
\end{figure*}

\subsection{Quality validation of the density reconstruction}

In order to show the performance of the Poisson--lognormal filter we compare the results with two other estimates of the density field. 
We follow \citet[][]{kitaura} and \citet[][]{kitaura_sdss} in quantitatively measuring the quality of the reconstructions.

\subsubsection{Alternative density field estimators: Inverse weighting and LSQ--Wiener filtering}

\label{sec:filters}

Let us first introduce a representation of the data which tries to compensate for the selection function effect which we call inverse weighting (IW).
We define the inverse weighted galaxy number count per cell $i$ as:
\begin{equation}
{N}_{{\rm g},i}^{\rm IW}\equiv\frac{1}{w_i} N^{\rm o}_{{\rm g},i}{.}
\label{def:IW}
\end{equation}
The corresponding inverse weighted overdensity is calculated as follows:
\begin{equation}
\delta_{{\rm g},i}^{\rm IW}\equiv \frac{N_{{\rm g},i}^{\rm IW}}{\overline{N}_{\rm g}}-1{.}
\label{eq:datamodel} 
\end{equation}
Note that the inverse weighting scheme can be derived as the maximum likelihood estimator assuming a Poissonian likelihood \citep[for a derivation see][]{kitaura_sdss}. As discussed in \citet[][]{kitaura_sdss} IW boosts the estimated density field at low completeness. Therefore it includes in general an additional smoothing step which lessons this effect  \citep[see e.g.~][]{2004MNRAS.352..939E}.

For an additional comparison let us introduce the least squares version of the Wiener filter (or LSQ filter for short) given by \citet[see][]{kitaura_sdss}:
\begin{equation}
{\mbi \delta}_{\rm M}^{\rm LSQ}\equiv\mat B^{-1}\left(\mat S^{-1}+\mat W^{\dagger}\mat N^{-1}\mat W\right)^{-1}\mat W^{\dagger}\mat N^{-1}{\mbi \delta}_{{\rm g}}^{{\rm o}}{,}
\label{eq:WF}
\end{equation}
with $\mat W$ being the three dimensional mask operator defined by:
$W_{i,j}\equiv w_j \delta^{\rm K}_{i,j}$ ($\delta^{\rm K}_{i,j}$ is the Kroenecker delta), and the Fourier transform of
$\mat B$ given by: $\hat{\hat{B}}_{k,k'}\equiv b_{k'}\delta^{\rm
K}_{k,k'}$ as introduced in subsection \ref{sec:linbias}. 
We define the observed galaxy overdensity which we use as the input vector for the LSQ reconstruction by:
\begin{equation}
{\delta}^{\rm o}_{{\rm g},i}\equiv \frac{N^{\rm o}_{{\rm g},i}}{\overline{N}_{\rm g}}-w_i{.}
\label{eq:datamodel} 
\end{equation}
The noise term in Eq.~\ref{eq:WF} has the following form \citep[see][]{kitaura_sdss}:
\begin{equation}
N_{i,j}\equiv\frac{w_i}{\overline{N}_{\rm g}}\delta^{\rm K}_{i,j}{.}
\label{eq:noiseav}
\end{equation}
Note that the LSQ--Wiener filter  is the optimal linear filter using up to second order statistics even for the Poisson--noise assumption for which the noise is signal dependent. 
There is not an additional assumption or approximation by neglecting the signal to noise correlation. This point has been unclear in the literature \citep[see for example][]{1995ApJ...449..446Z,1998ApJ...503..492S,2004MNRAS.352..939E}. We show that the signal and the noise are indeed uncorrelated in the appendix.  The LSQ--Wiener filter also happens to be the MAP filter for a Gaussian likelihood and a Gaussian prior as indicated in table \ref{tab:filters}.

In our numerical experiments we use a unity bias:
$b_{k}=1$. Thus, the Fourier transform of $\mat S$ yields:
$P_{{\rm g}}(\mbi k)=P_{{\rm M}}(\mbi k)$.  The power--spectrum $P_{\rm M}(\mbi k)$ is
given by a nonlinear fit that also describes the effects of virialised
structures with a halo term as given by
\citet[][]{2003MNRAS.341.1311S} at redshift $z=0$. We choose the
concordance $\Lambda$CDM--cosmology with $\Omega_{\rm m}=0.24$,
$\Omega_{\rm K}=0$ and $\Omega_{\Lambda}=0.76$ \citep[][]{spergel}.
In addition, we assumed a Hubble constant with $h=73$ and a spectral
index $n_s=1$.

\subsubsection{Quantitative measures}

Let us  define the correlation coefficient $\rm r$ between the reconstructed and original matter density fields by:
\begin{equation}
{\rm r}(\delta^{\rm rec},\delta_{\rm M})\equiv\frac{\sum^{N_{\rm cells}}_i\delta_{{\rm M},i}\delta^{\rm rec}_i}{\sqrt{\sum^{N_{\rm cells}}_i\left(\delta_{{\rm M},i}\right)^2}\sqrt{\sum^{N_{\rm cells}}_j\left(\delta^{\rm rec}_j\right)^2}}{,}
\end{equation}
and the Euclidean distance
\begin{equation}
{\rm D}_{\rm Euc}(\delta^{\rm rec},\delta_{\rm M})\equiv\sqrt{\frac{1}{N_{\rm cells}} \sum^{N_{\rm cells}}_i\,\left(\delta^{\rm rec}_i-\delta_{{\rm M},i})\right)^2}{.}
\label{eq:enseres}
\end{equation}

\subsection{Input data setup}

\label{sec:input}

 We construct the mock  observed galaxy distribution taking a random subsample of the particles in  the Millennium run at redshift zero \citep[see][]{2005Natur.435..629S} which was gridded on a $256^3$ mesh. Later we also investigate the resolution dependence using a $128^3$, and a $64^3$ mesh.  As already stated above, our setup permits us to avoid the galaxy biasing problem in our tests. 
Note, that we also avoid the redshift distortions by considering the dark matter particles in real--space.
In this way we generate three different  input mock galaxy catalogues. One  has about 1 Million particles and has been produced as a Poisson sampling with a homogeneous completeness of  $w=10^{-4}$. The other two mocks were generated with a radial selection function using two exponential decaying models of completeness $w$ (see Fig.~\ref{fig:sel}) emulating apparent magnitude limit effects \citep[see radial selection function in][]{kitaura_sdss}. The final mock galaxy samples have 350961 and 123679 particles using the softer and steeper decaying selection functions respectively. The observer was set at the center of the box, i.e.~at coordinates: X=250 Mpc/h, Y=250 Mpc/h, and Z=250 Mpc/h.

The left panel on Fig.~\ref{fig:corr_gal} shows a cell--to--cell comparison between the dark matter distribution of the Millennium run ($\sim10^{10}$ particles) gridded on a $256^3$ mesh and a subsample of 1 Million homogeneously selected mock galaxies of the \citet{DeLucia-Blaizot-07} catalogue.
One can clearly see a deviation of the pixels with respect to the perfect slope of $45^\circ$. This effect is due to galaxy biasing. 
 The right panel shows the analougous comparison with a Poisson sampling using a homogeneous completeness of $w=10^{-4}$ which leaves  about 1 Million particles.
Here, we see  a nearly perfect scatter around the $45^\circ$ slope demonstrating that our mocks do not include biasing. 

\subsection{MAP results}

Here we calculate the maximum a posteriori solutions which we derived in the previous theoretical sections. There we assumed two different prior distributions for the matter field: a Gaussian and a lognormal prior.

\subsubsection{Gaussian prior and Poissonian likelihood}

In this subsection we present the results given by the maximum a posteriori solution assuming a Gaussian prior and a Poissonian likelihood.
The solution of Eqn.~\ref{eq:MAPG} leads to a matter field which dramatically underestimates large overdensities (see Fig.~\ref{fig:gapmap}). This shows that the Gaussian prior cannot fit the underlying matter field which has a clearly non--Gaussian distribution with a minimum overdensity of $\sim-1$ up to maximal overdensities of about 1500 at the resolution we are looking at ($\sim2$ Mpc/h cell side length). The density peaks are highly suppressed with a Gaussian prior. This effect is known from the  Wiener filter as traditionally applied where the noise covariance is dependent on the signal \citep[see discussion in][]{kitaura_sdss}. Note, that the filter we are using here is a more accurate being based on the full Poissonian distribution and not only on the second order term as in the Wiener filter.

\subsubsection{Lognormal prior and Poissonian likelihood}

Here we present the results of the maximum a posteriori solution assuming a lognormal prior and a Poissonian likelihood. For that we solve the MAP Eqn.~\ref{eq:MAPLNUB}.

\begin{figure*}
\begin{tabular}{cc}
\includegraphics[width=8.5cm]{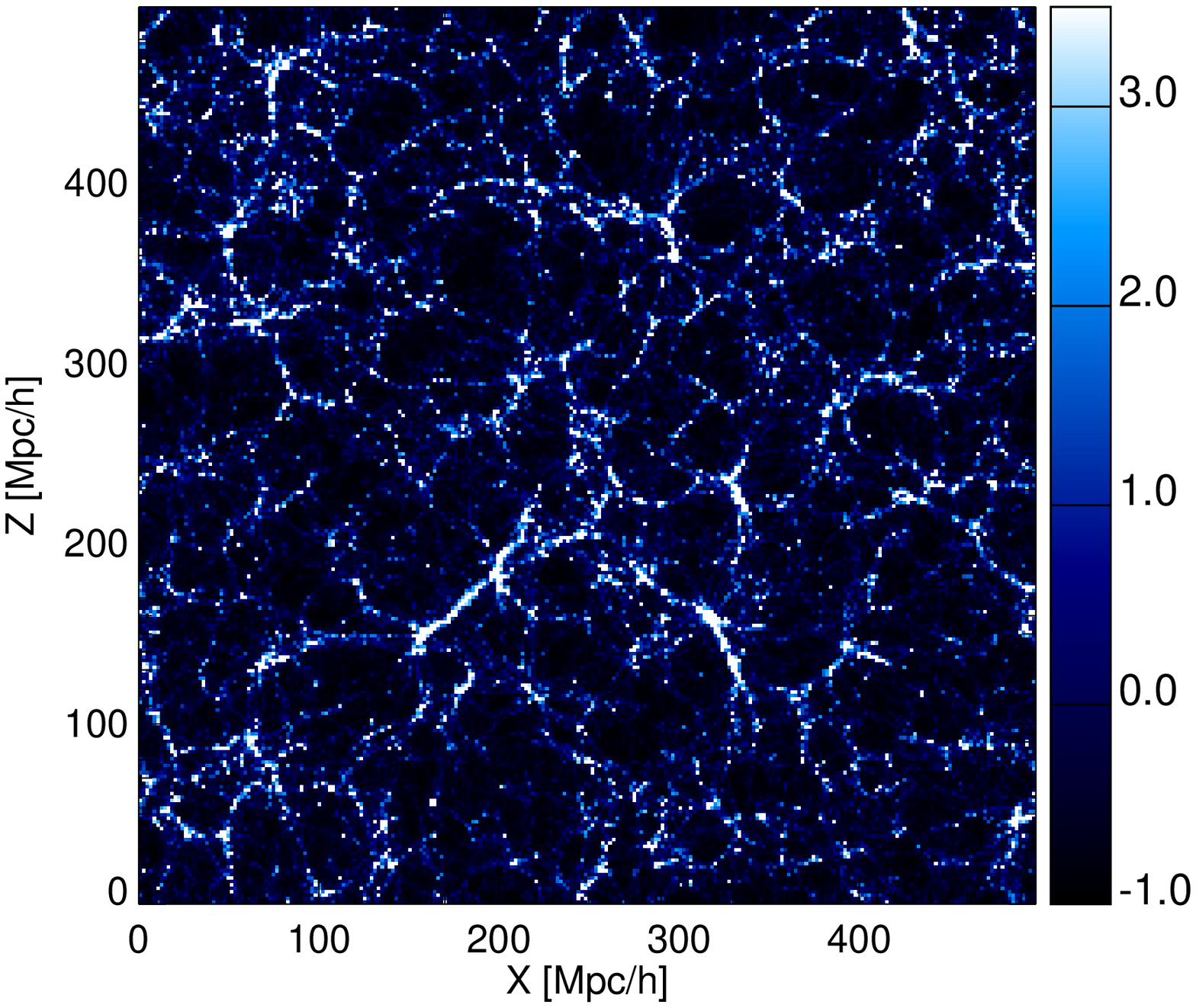}
\put(-230,0.5){{\huge (a)}}
\put(-5,105.){\rotatebox[]{0}{$\delta_{\rm M}$}}
\hspace{0.5cm}
\includegraphics[width=8.5cm]{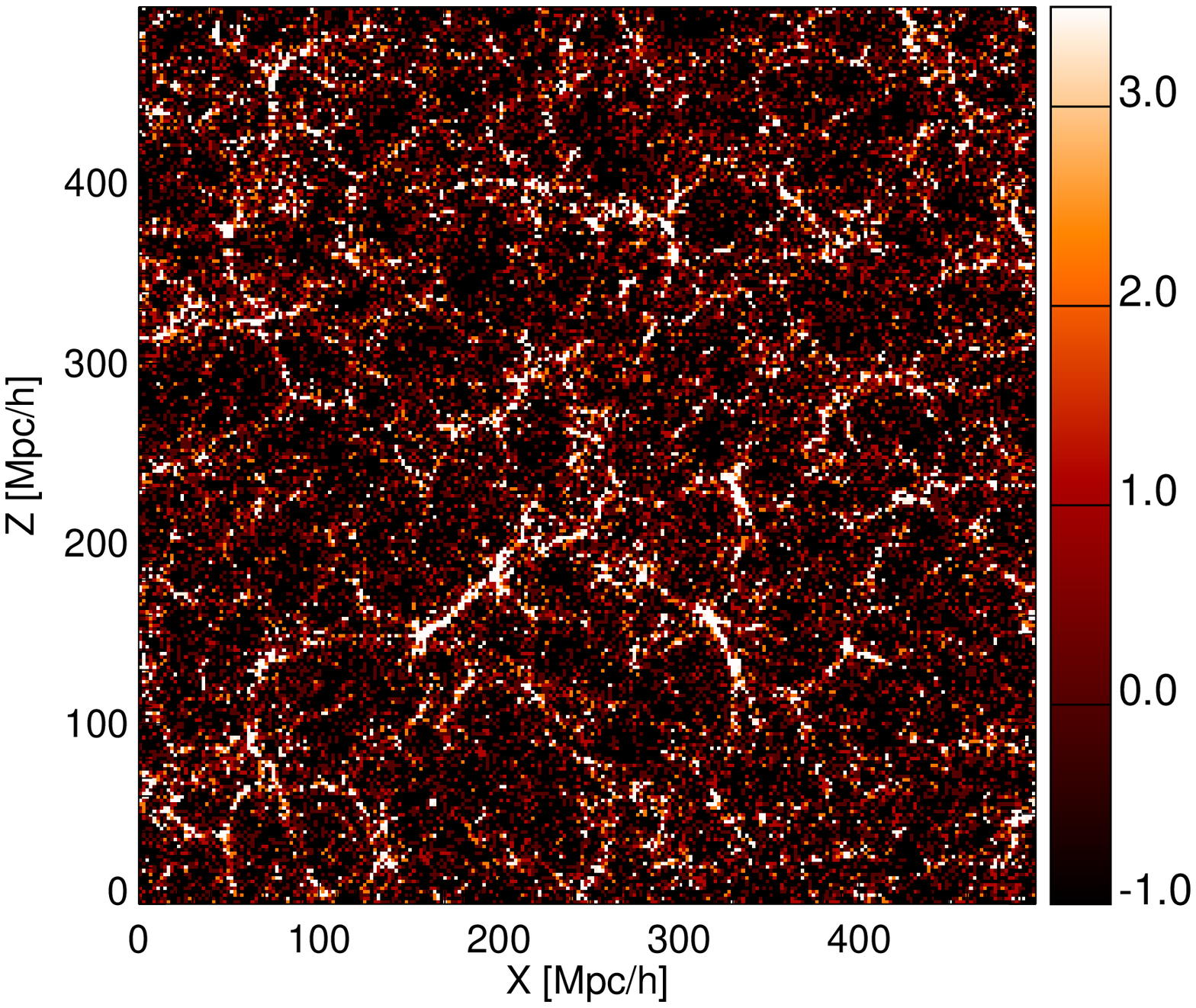}
\put(-230,1.5){{\huge (b)}}
\put(-5,105.){\rotatebox[]{0}{$\delta_{\rm g}$}}
\\
\includegraphics[width=8.5cm]{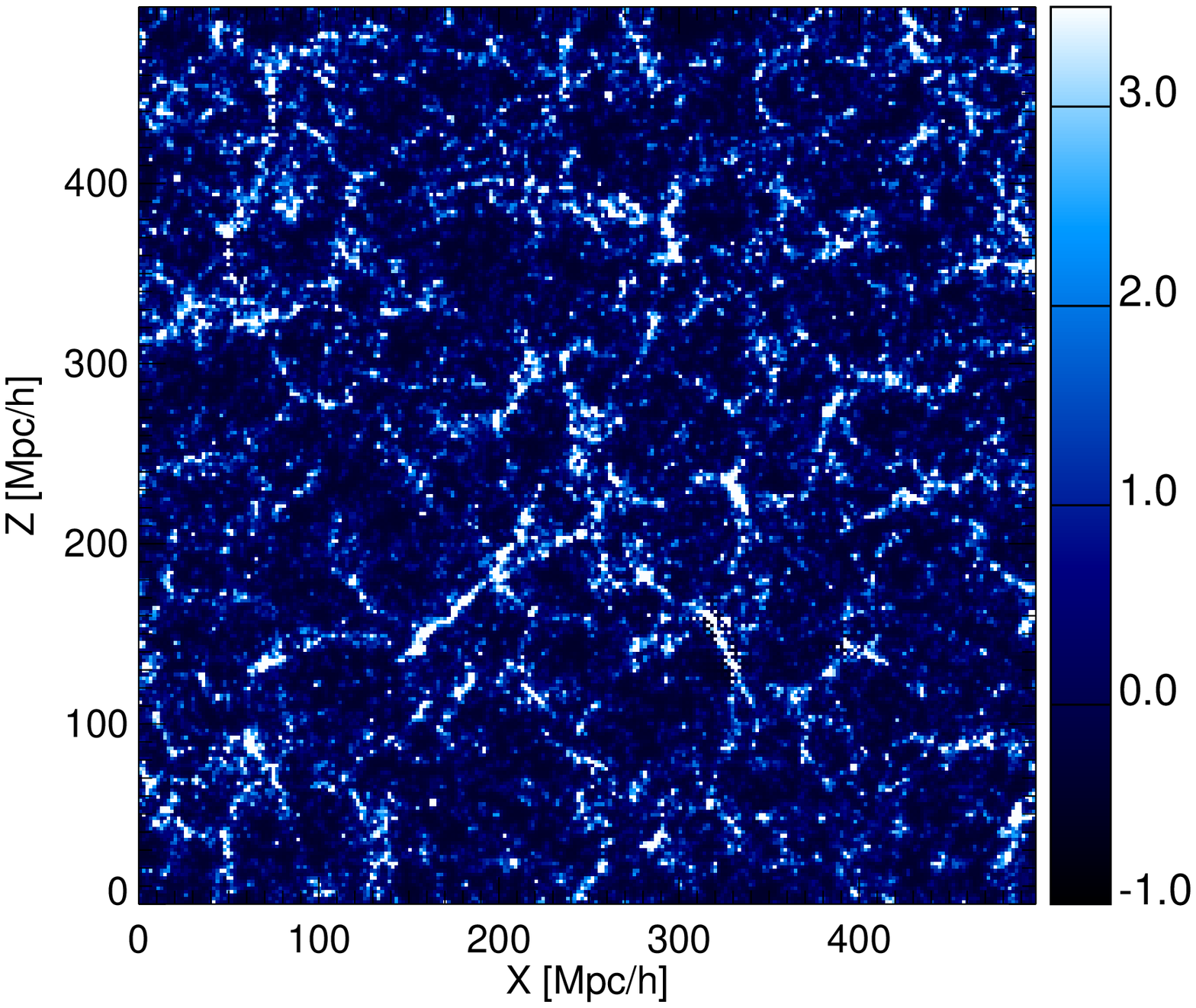}
\put(-230,0.5){{\huge (c)}}
\put(-5,105.){\rotatebox[]{0}{$\delta_{\rm M}^{\rm L}$}}
\hspace{0.5cm}
\includegraphics[width=8.5cm]{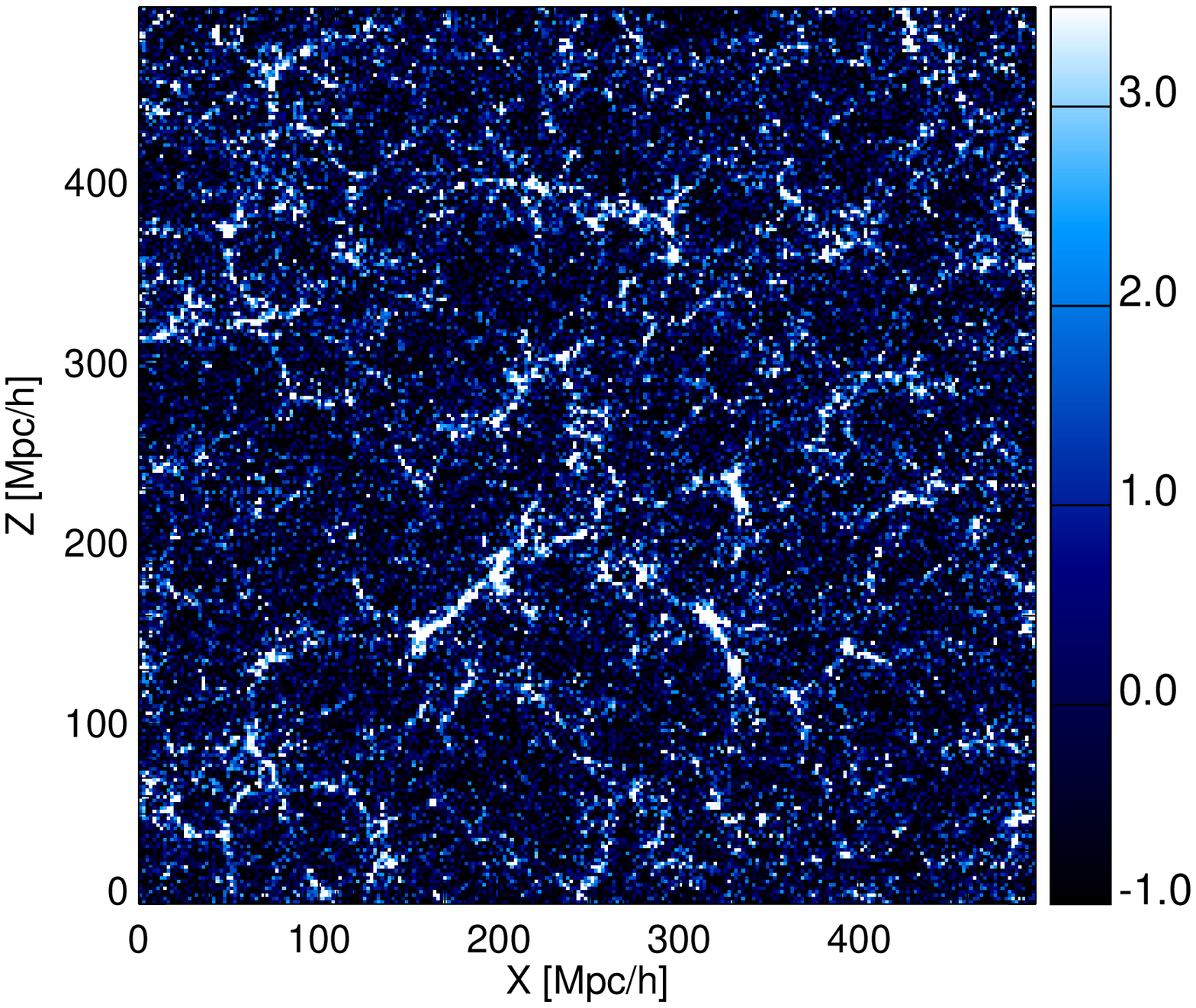}
\put(-230,1.5){{\huge (d)}}
\put(-5,105.){\rotatebox[]{0}{$\delta_{\rm M}^{\rm LSQ}$}}
\\
\includegraphics[width=6.cm]{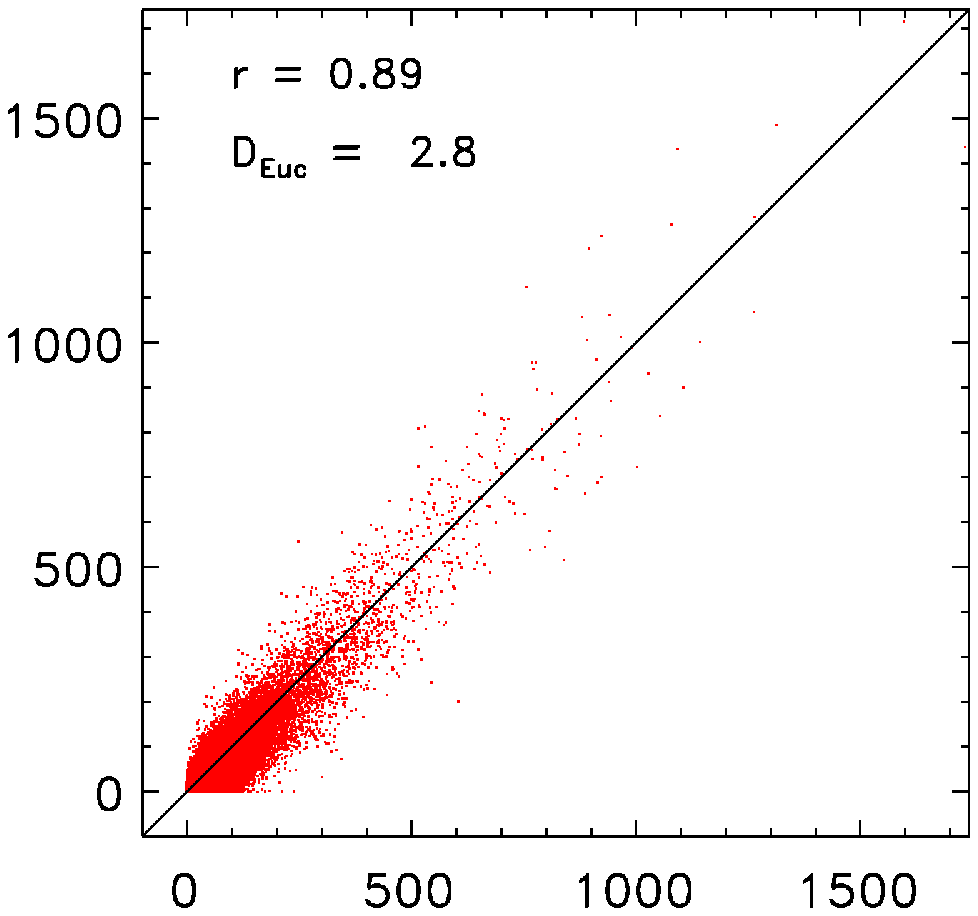}
\put(-185,0.5){{\huge (e)}}
\put(-50,160){\rotatebox[]{0}{\large$256^3$}}
\put(-185,100.){\rotatebox[]{90}{$\delta_{\rm M}^{\rm L}$}}
\put(-80,10){\rotatebox[]{0}{\large$\delta_{\rm M}$}}\hspace{0.5cm}
\hspace{2.5cm}
\includegraphics[width=6.cm]{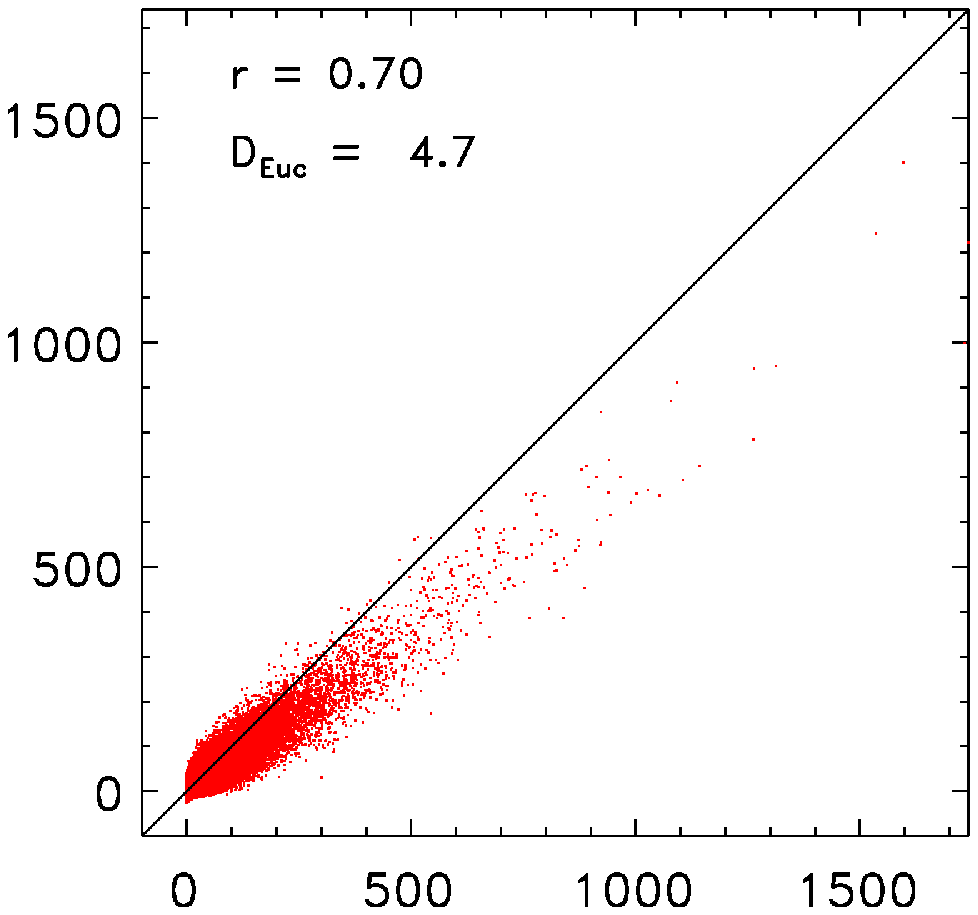}
\put(-185,1.5){{\huge (f)}}
\put(-50,160){\rotatebox[]{0}{\large$256^3$}}
\put(-185,100.){\rotatebox[]{90}{$\delta_{\rm M}^{\rm LSQ}$}}
\put(-80,10){\rotatebox[]{0}{\large$\delta_{\rm M}$}}\hspace{0.5cm}
\end{tabular}
\caption{  Panel a: slice through the Millenium run dark matter particle simulation.   Panel b: mock galaxy distribution with $10^6$ particles. Panel c: reconstruction with the lognormal filter. Panel d: reconstruction with the LSQ-Wiener filter. Panel e: cell--to--cell correlation between the overdensity of the full three--dimensional reconstruction with the Lognormal filter (panel c) and the matter field (panel a).  Panel f: cell--to--cell correlation between the  full three--dimensional overdensity of the reconstruction with the LSQ-Wiener filter (panel d) and the matter field (panel a). The cell--to--cell correlation between the mock galaxy distribution (panel b) and the dark matter distribution can be seen on the right panel of Fig.~\ref{fig:corr_gal}. The plots were produced by calculating the mean over 15 neighboring slices around slice 218 (Y$\sim176$ Mpc/h) through a 500 Mpc/h cube box with a $256^3$ grid.} 
\label{fig:rec_MRUS}
\end{figure*}

\begin{figure*}
\begin{tabular}{cc}
\includegraphics[width=8.5cm]{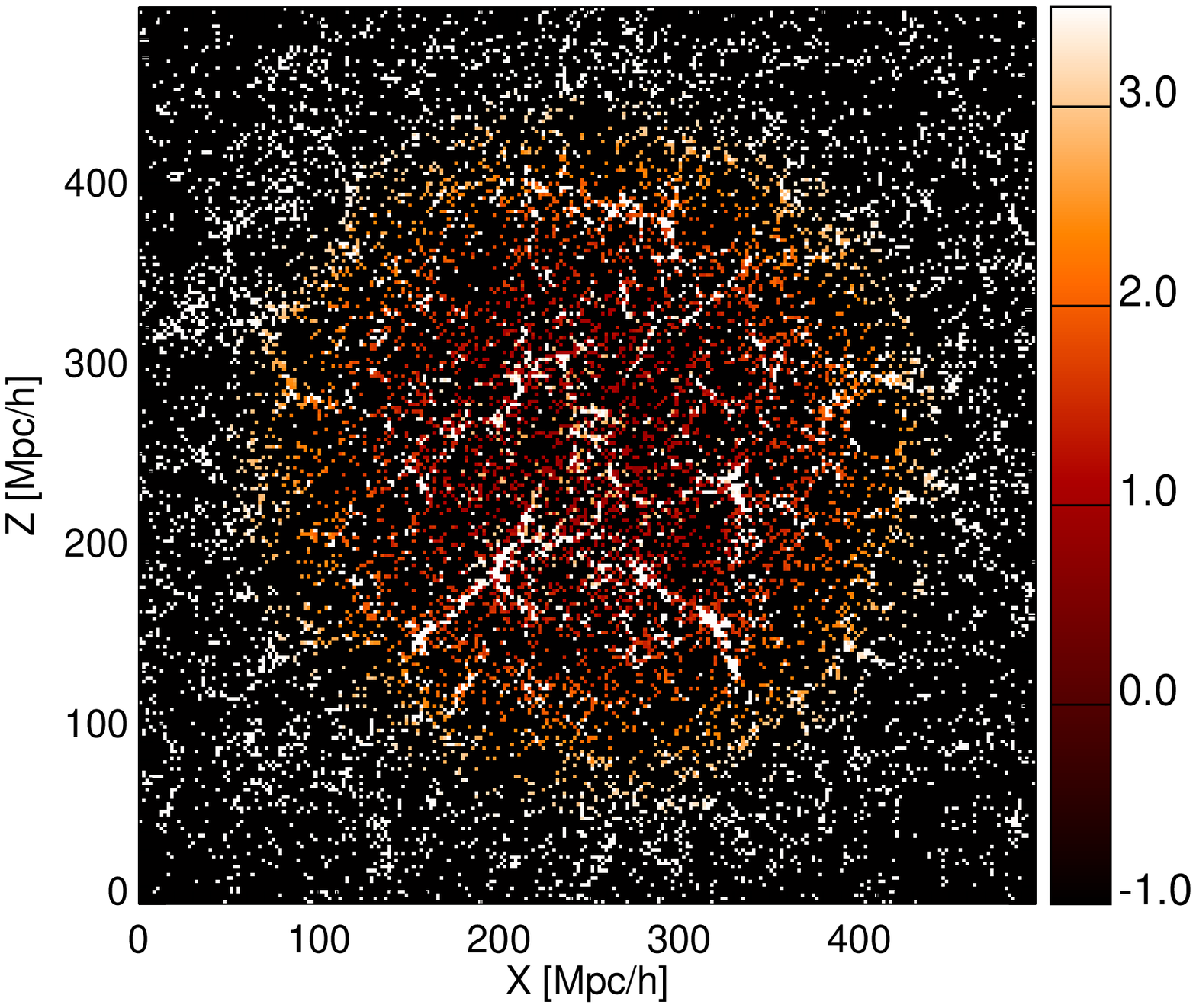}
\put(-230,0.5){{\huge (a)}}
\put(-5,105.){\rotatebox[]{0}{$\delta^{\rm IW}_{\rm g}$}}
\hspace{1.5cm}
\includegraphics[width=6.5cm]{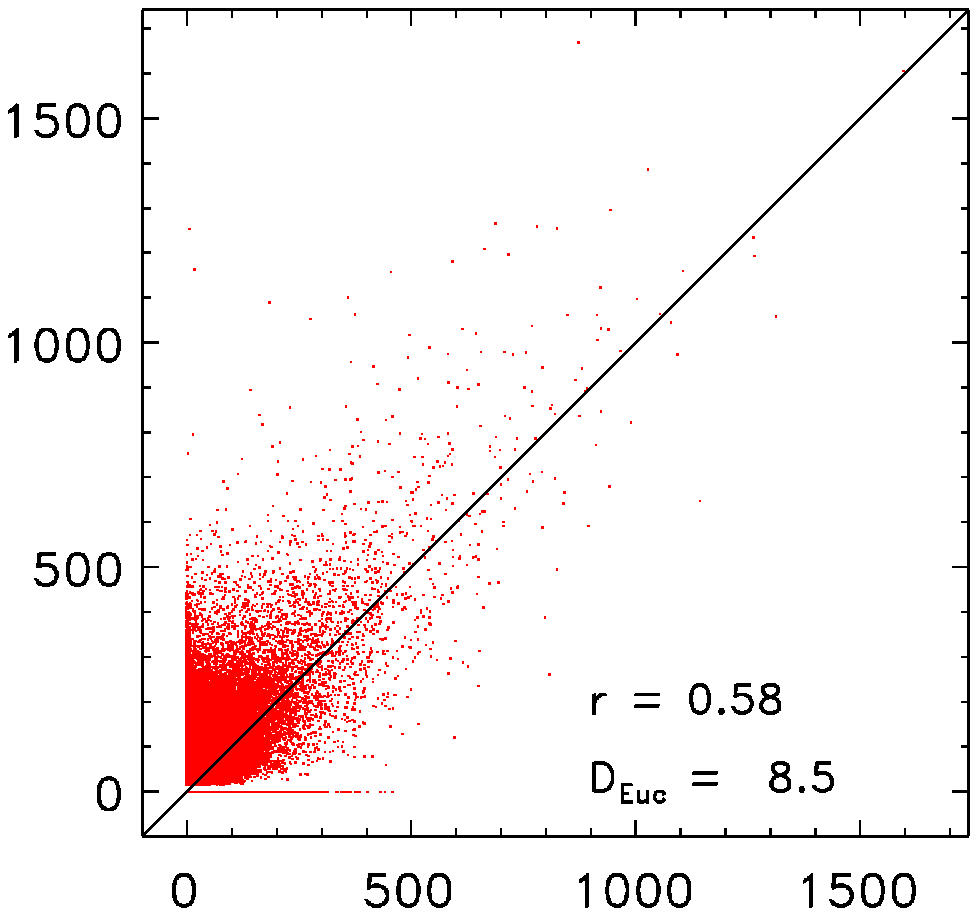}
\put(-185,0.5){{\huge (b)}}
\put(-50,170){\rotatebox[]{0}{\large$256^3$}}
\put(-50,180){\rotatebox[]{0}{\large$w_1$}}
\put(-200,105.){\rotatebox[]{90}{$\delta_{\rm g}^{\rm IW}$}}
\put(-85,10){\rotatebox[]{0}{\large$\delta_{\rm M}$}}
\\
\includegraphics[width=8.5cm]{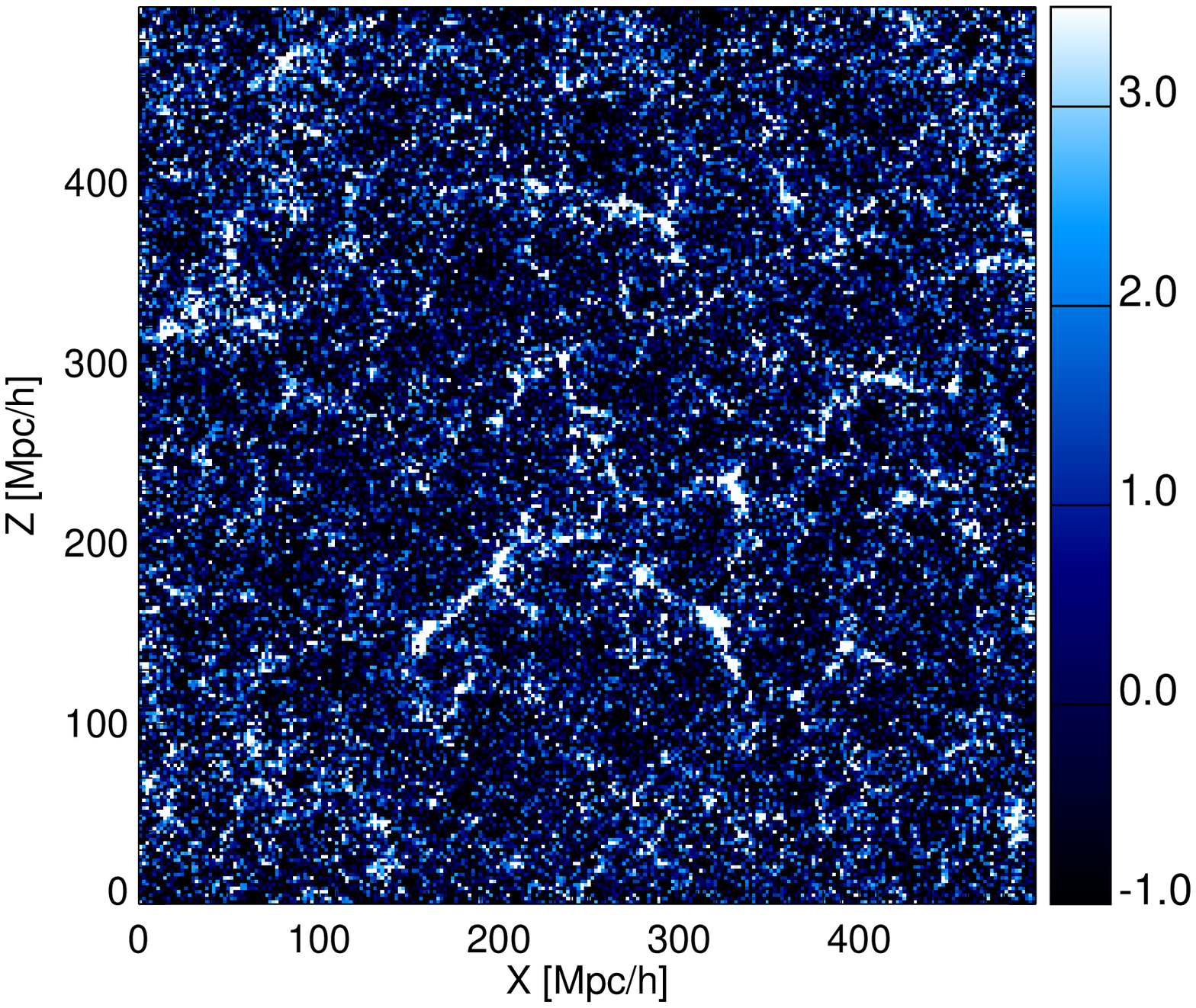}
\put(-230,0.5){{\huge (c)}}
\put(-5,105.){\rotatebox[]{0}{$\delta_{\rm M}^{\rm LSQ}$}}
\hspace{1.5cm}
\includegraphics[width=6.5cm]{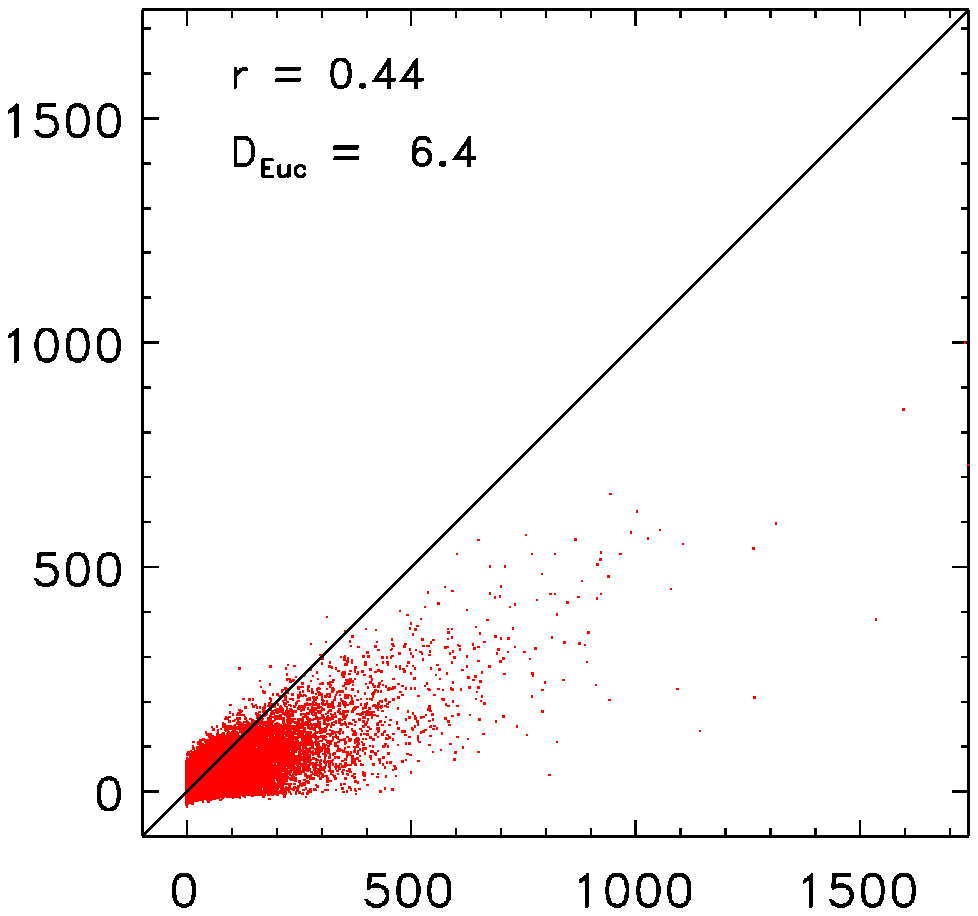}
\put(-185,0.5){{\huge (d)}}
\put(-50,170){\rotatebox[]{0}{\large$256^3$}}
\put(-50,180){\rotatebox[]{0}{\large$w_1$}}
\put(-200,105.){\rotatebox[]{90}{$\delta_{\rm M}^{\rm LSQ}$}}
\put(-85,10){\rotatebox[]{0}{\large$\delta_{\rm M}$}}
\\
\includegraphics[width=8.5cm]{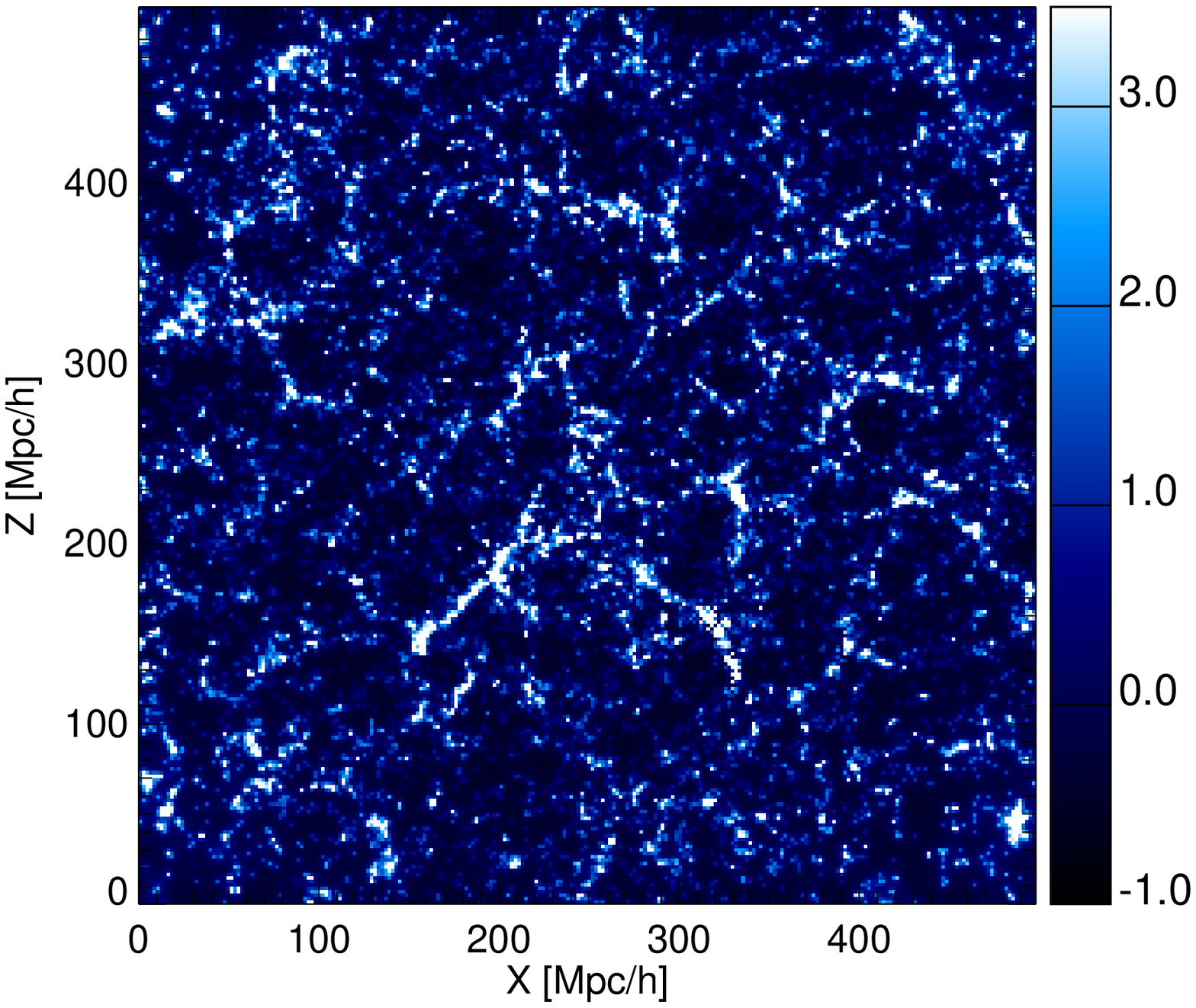}
\put(-230,0.5){{\huge (e)}}
\put(-5,105.){\rotatebox[]{0}{$\delta_{\rm M}^{\rm L}$}}
\hspace{1.5cm}
\includegraphics[width=6.5cm]{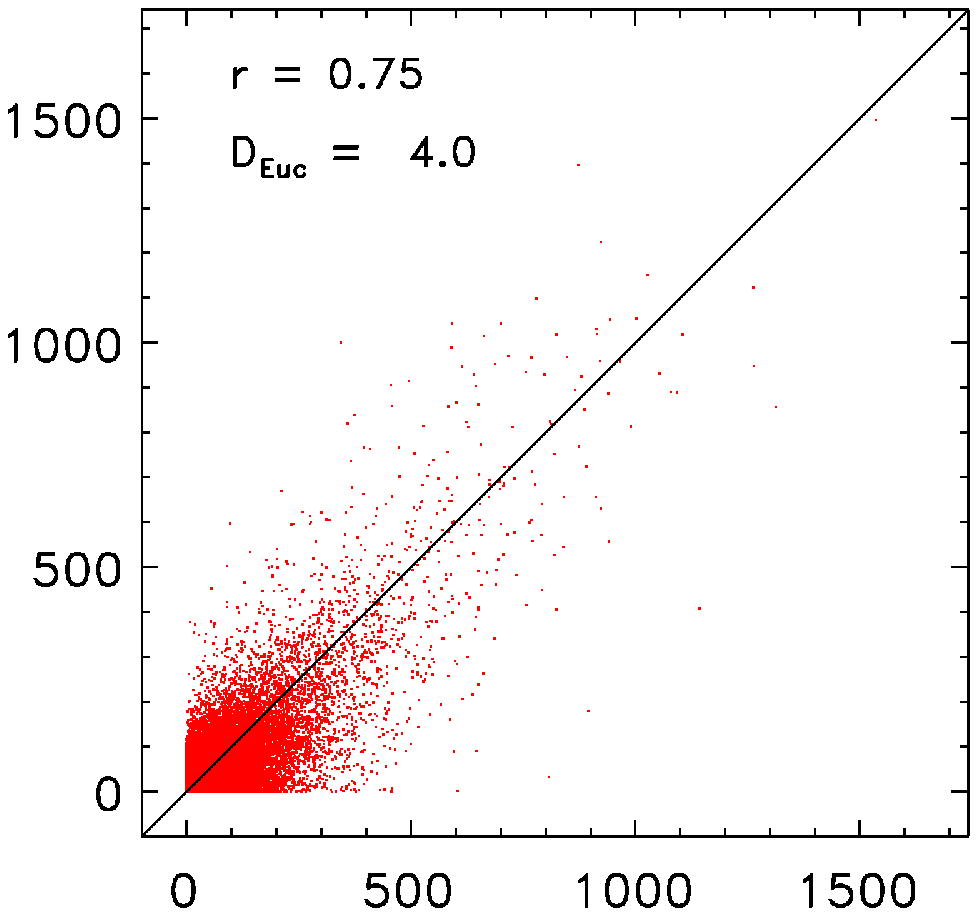}
\put(-185,0.5){{\huge (f)}}
\put(-50,170){\rotatebox[]{0}{\large$256^3$}}
\put(-50,180){\rotatebox[]{0}{\large$w_1$}}
\put(-200,105.){\rotatebox[]{90}{$\delta_{\rm M}^{\rm L}$}}
\put(-85,10){\rotatebox[]{0}{\large$\delta_{\rm M}$}}
\end{tabular}
\caption{ Panel a: inverse weigted mock galaxy distribution after applying a radial selection function ($w_1$) leaving 350961 galaxies. Panels c: LSQ-Wiener filter reconstruction. Panels e: Lognormal filter reconstruction. The plots were produced by calculating the mean over 15 neighboring slices around slice 218 (Y$\sim176$ Mpc/h) through a 500 Mpc/h cube box with a $256^3$ grid. The performance depicted in cell--to--cell correlation plots are shown in the right hand side panels b, e and f.} 
\label{fig:rec_MRsel}
\end{figure*}

\begin{figure*}
\begin{tabular}{cc}
\includegraphics[width=8.5cm]{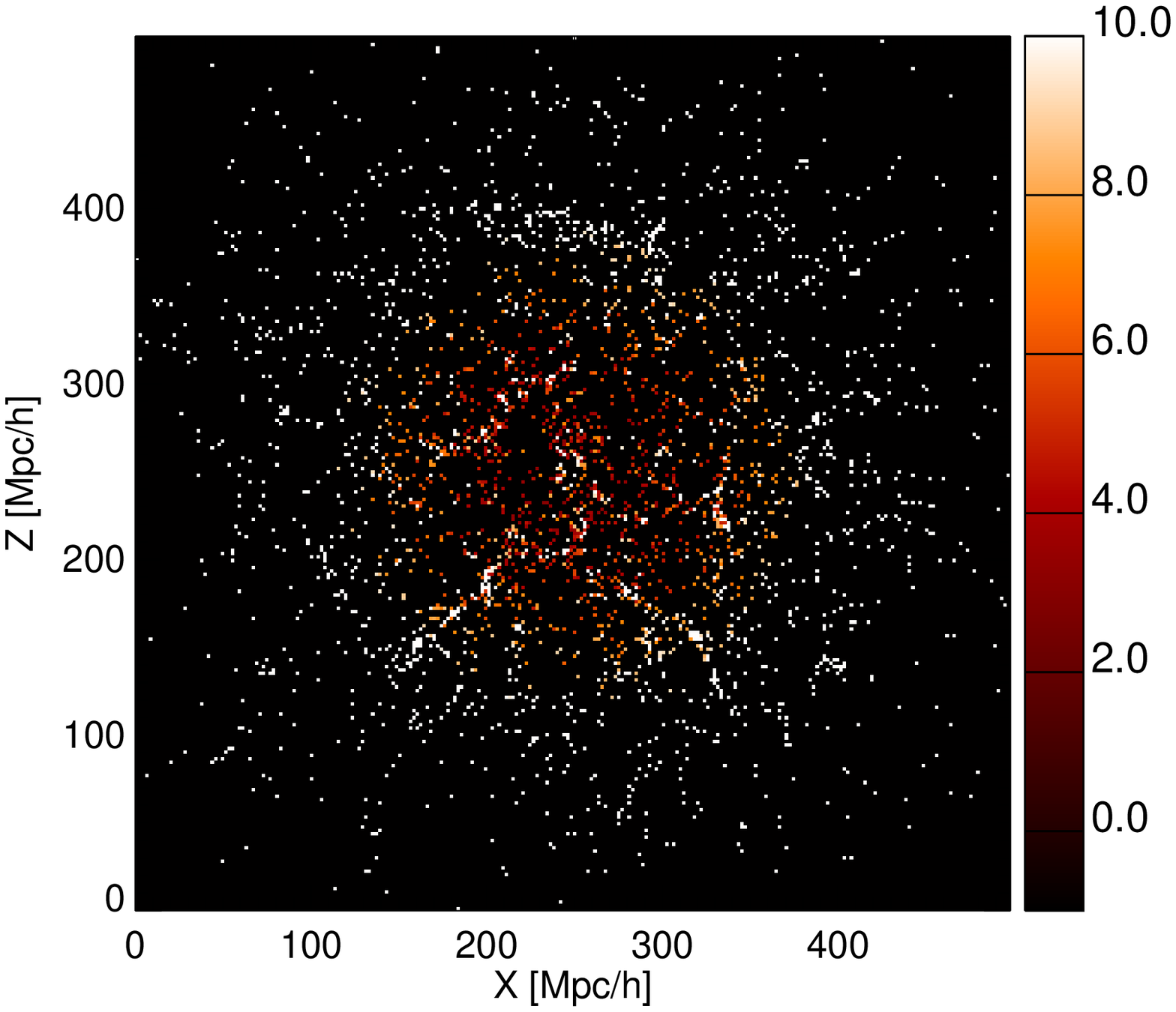}
\put(-230,0.5){{\huge (a)}}
\put(-5,105.){\rotatebox[]{0}{$\delta^{\rm IW}_{\rm g}$}}
\hspace{1.5cm}
\includegraphics[width=6.5cm]{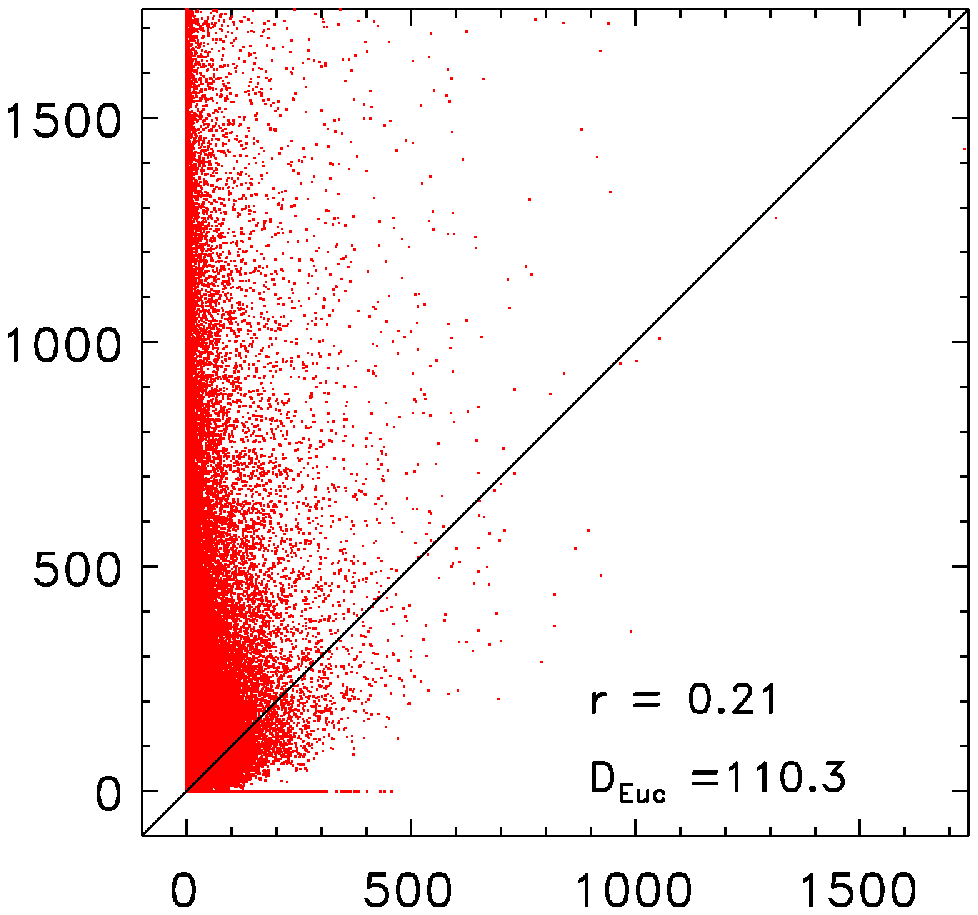}
\put(-185,0.5){{\huge (b)}}
\put(-50,170){\rotatebox[]{0}{\large$256^3$}}
\put(-50,180){\rotatebox[]{0}{\large$w_2$}}
\put(-200,105.){\rotatebox[]{90}{$\delta_{\rm g}^{\rm IW}$}}
\put(-85,10){\rotatebox[]{0}{\large$\delta_{\rm M}$}}
\\
\includegraphics[width=8.5cm]{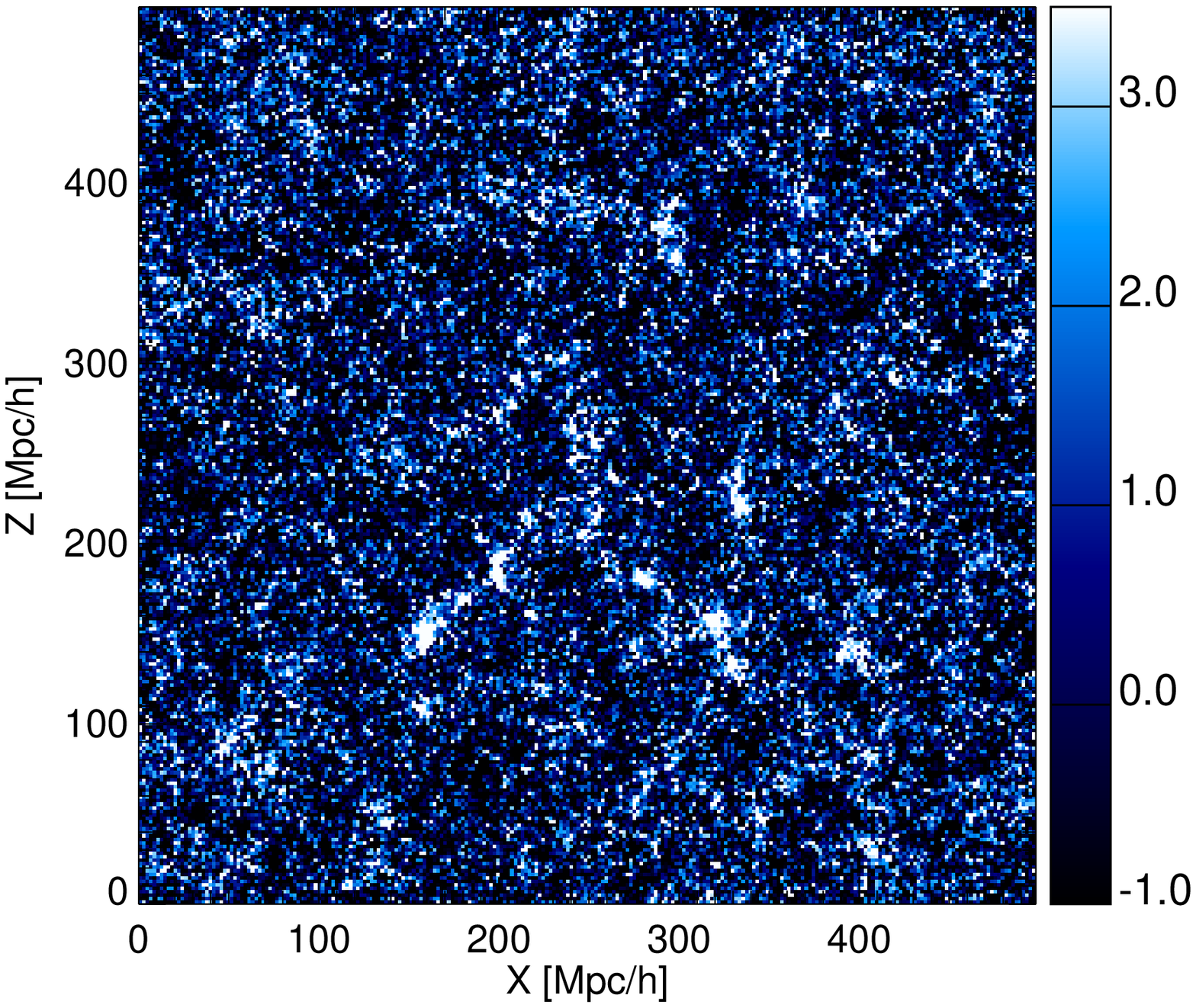}
\put(-230,0.5){{\huge (c)}}
\put(-5,105.){\rotatebox[]{0}{$\delta_{\rm M}^{\rm LSQ}$}}
\hspace{1.5cm}
\includegraphics[width=6.5cm]{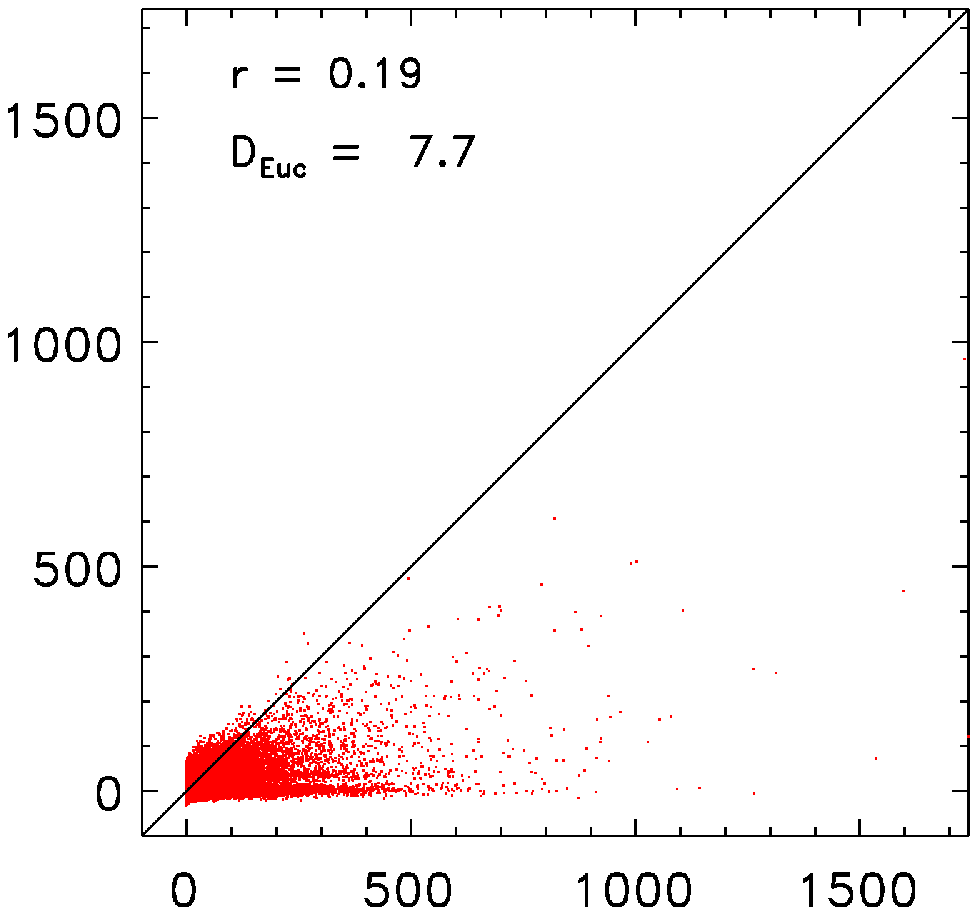}
\put(-185,0.5){{\huge (d)}}
\put(-50,170){\rotatebox[]{0}{\large$256^3$}}
\put(-50,180){\rotatebox[]{0}{\large$w_2$}}
\put(-200,105.){\rotatebox[]{90}{$\delta_{\rm M}^{\rm LSQ}$}}
\put(-85,10){\rotatebox[]{0}{\large$\delta_{\rm M}$}}
\\
\includegraphics[width=8.5cm]{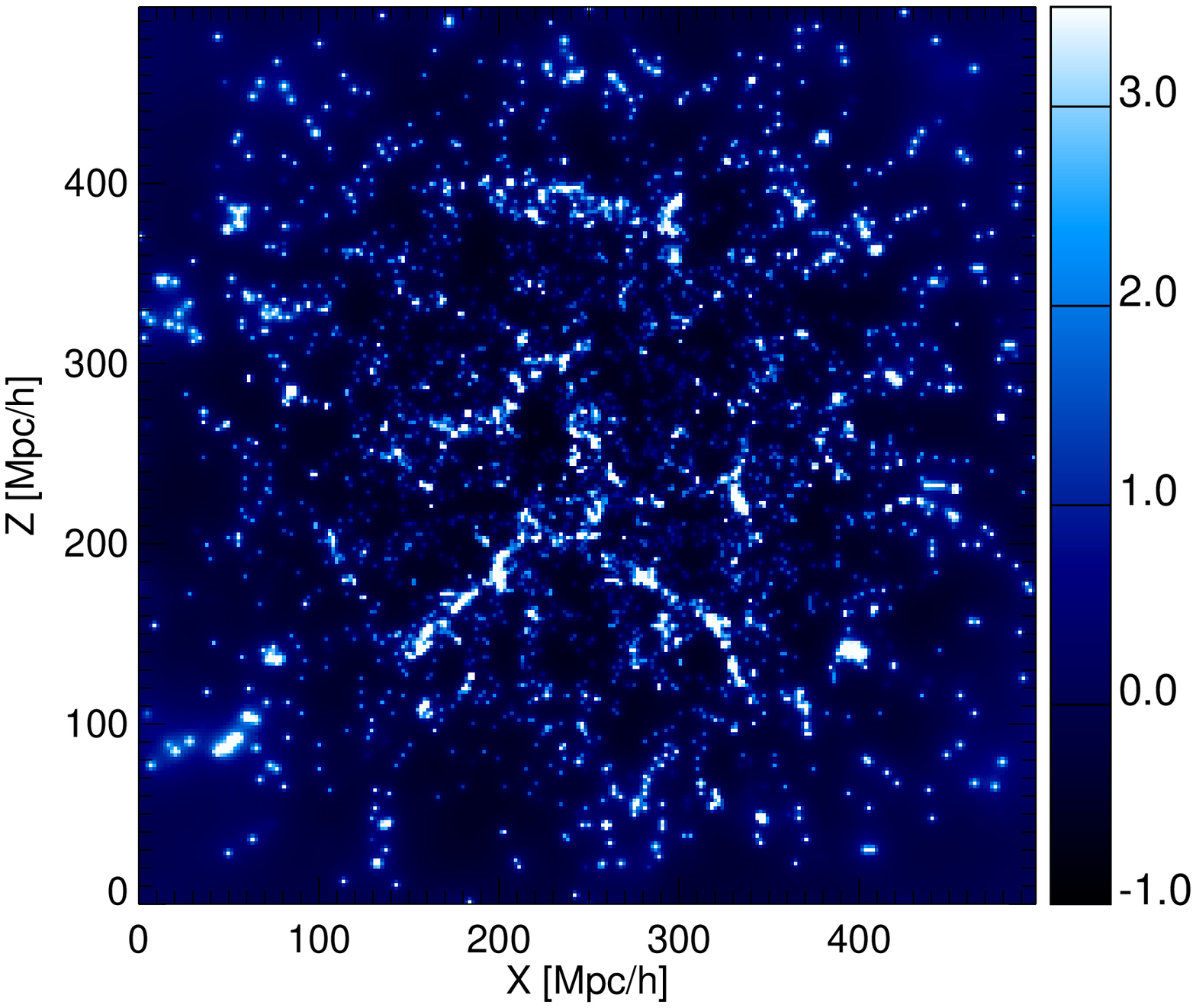}
\put(-230,0.5){{\huge (e)}}
\put(-5,105.){\rotatebox[]{0}{$\delta_{\rm M}^{\rm L}$}}
\hspace{1.5cm}
\includegraphics[width=6.5cm]{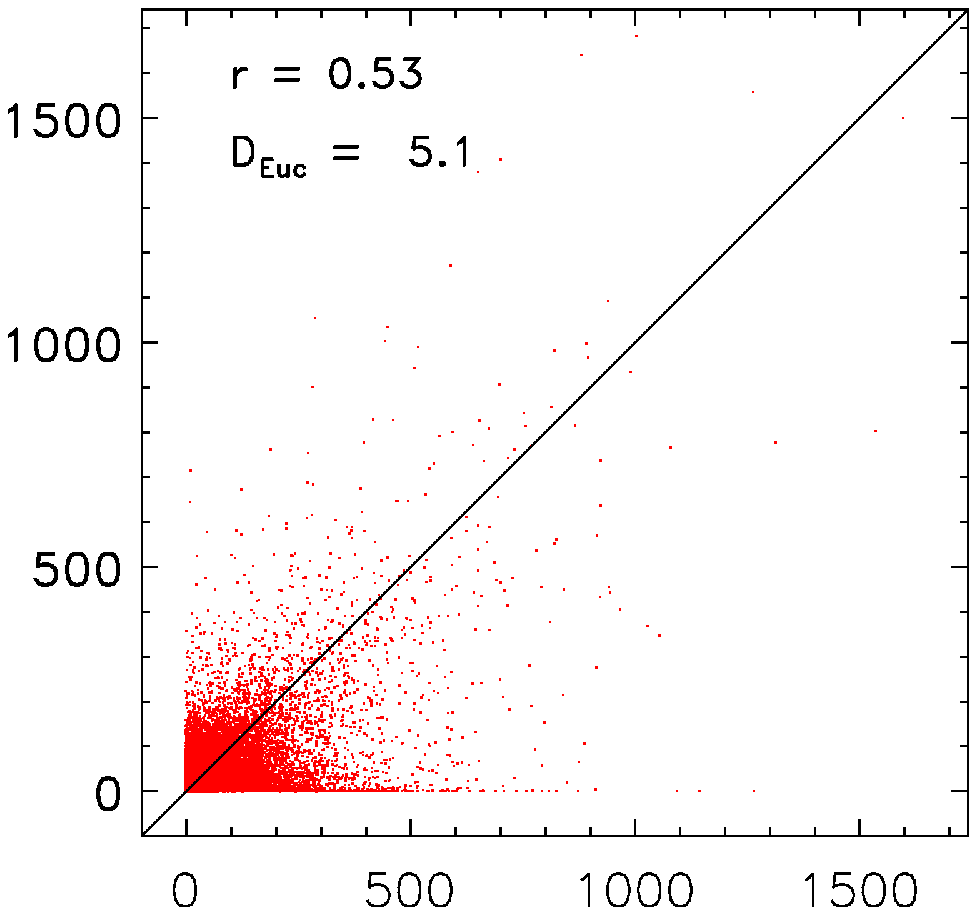}
\put(-185,0.5){{\huge (f)}}
\put(-50,170){\rotatebox[]{0}{\large$256^3$}}
\put(-50,180){\rotatebox[]{0}{\large$w_2$}}
\put(-200,105.){\rotatebox[]{90}{$\delta_{\rm M}^{\rm L}$}}
\put(-85,10){\rotatebox[]{0}{\large$\delta_{\rm M}$}}
\end{tabular}
\caption{ The same as Fig.~\ref{fig:rec_MRsel} corresponding to the radial selection function $w_2$ with a mock galaxy distribution of 123679 particles. } 
\label{fig:rec_MRsel2}
\end{figure*}

\begin{figure*}
\begin{tabular}{cc}
\includegraphics[width=8.5cm]{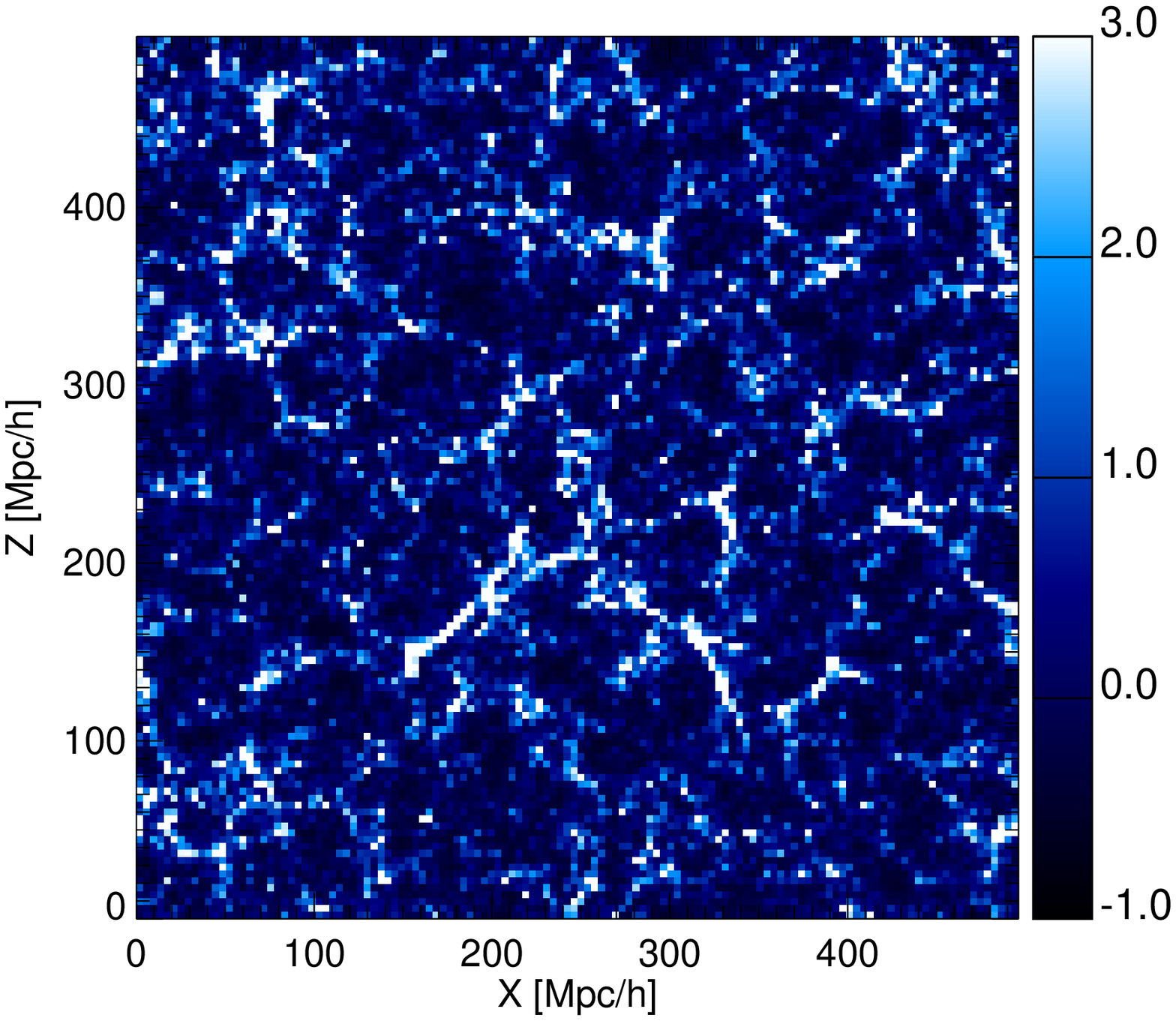}
\put(-230,0.5){{\huge (a)}}
\put(-5,105.){\rotatebox[]{0}{$\delta_{\rm M}^{\rm L}$}}
\hspace{0.5cm}
\includegraphics[width=8.5cm]{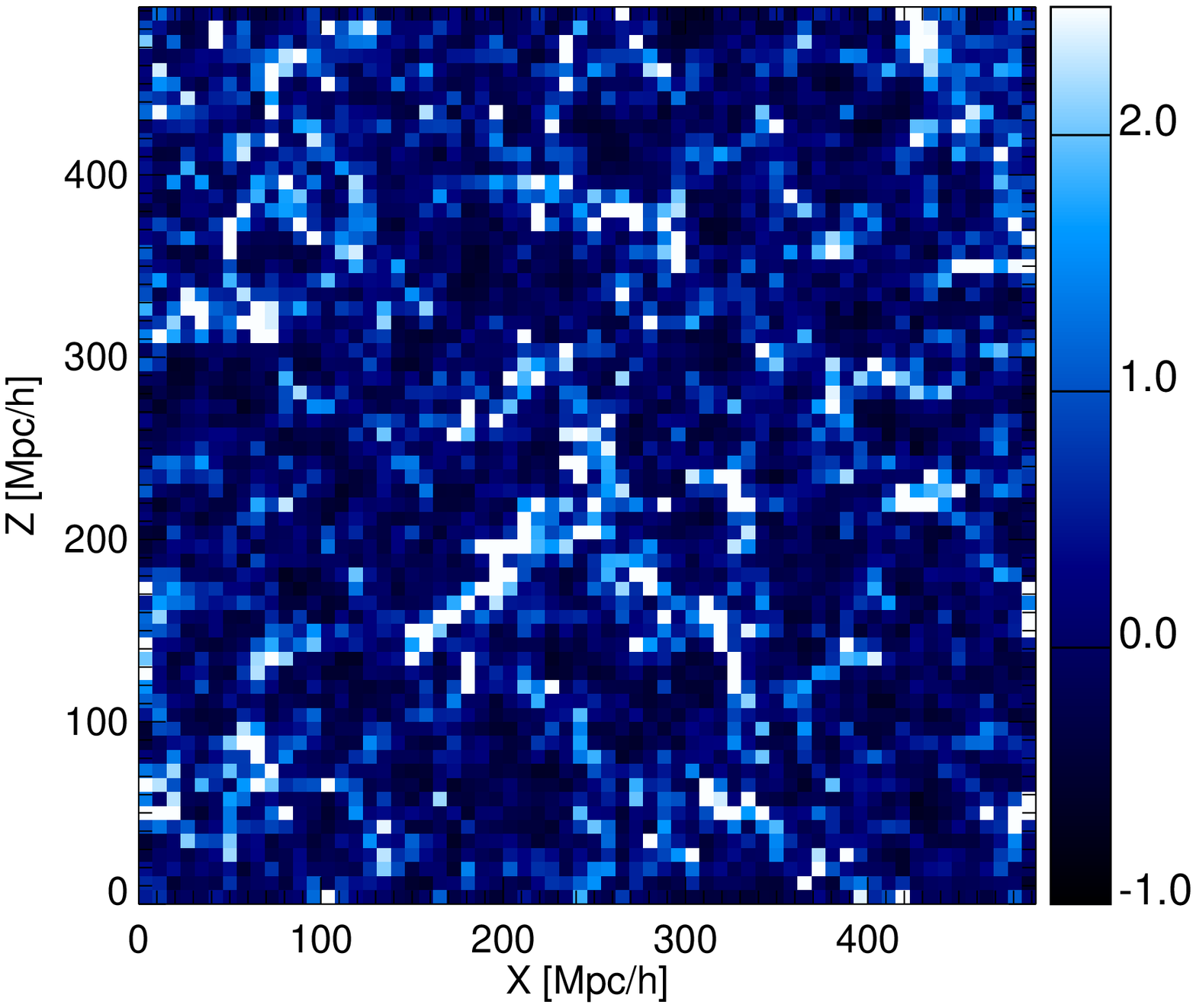}
\put(-230,1.5){{\huge (b)}}
\put(-5,105.){\rotatebox[]{0}{$\delta_{\rm M}^{\rm L}$}}
\\
\includegraphics[width=6.cm]{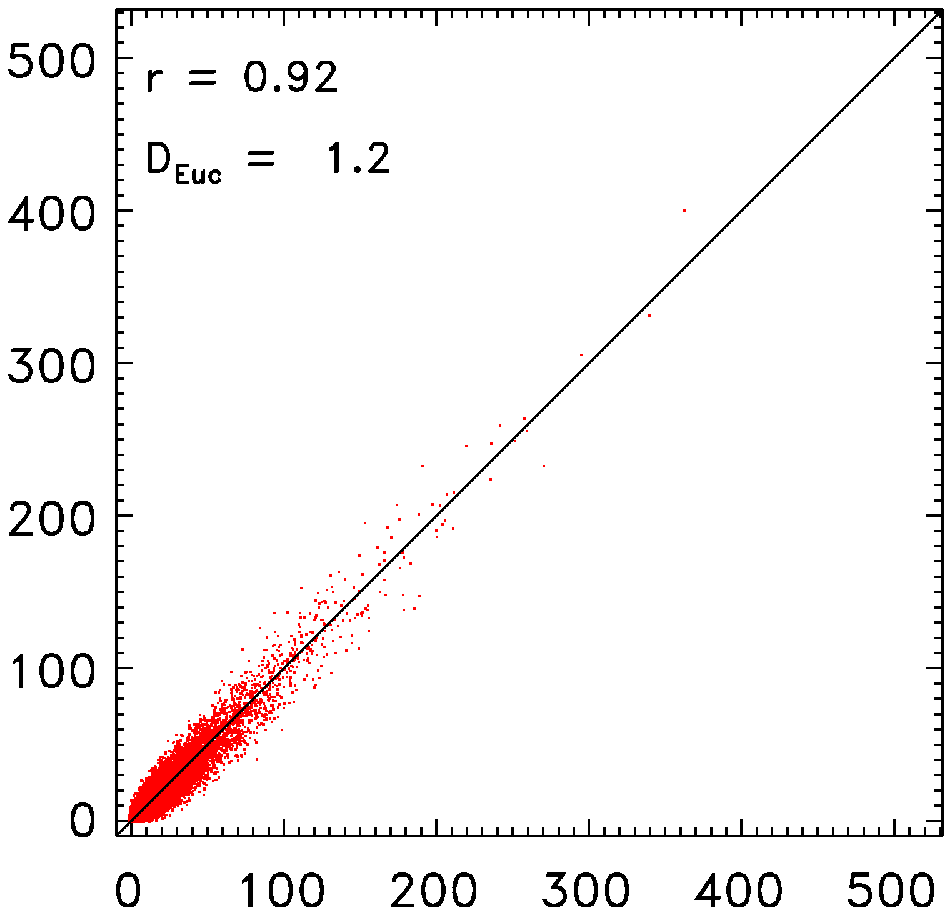}
\put(-185,0.5){{\huge (c)}}
\put(-50,160){\rotatebox[]{0}{\large$128^3$}}
\put(-185,100.){\rotatebox[]{90}{$\delta_{\rm M}^{\rm L}$}}
\put(-80,10){\rotatebox[]{0}{\large$\delta_{\rm M}$}}\hspace{0.5cm}
\hspace{2.5cm}
\includegraphics[width=6.cm]{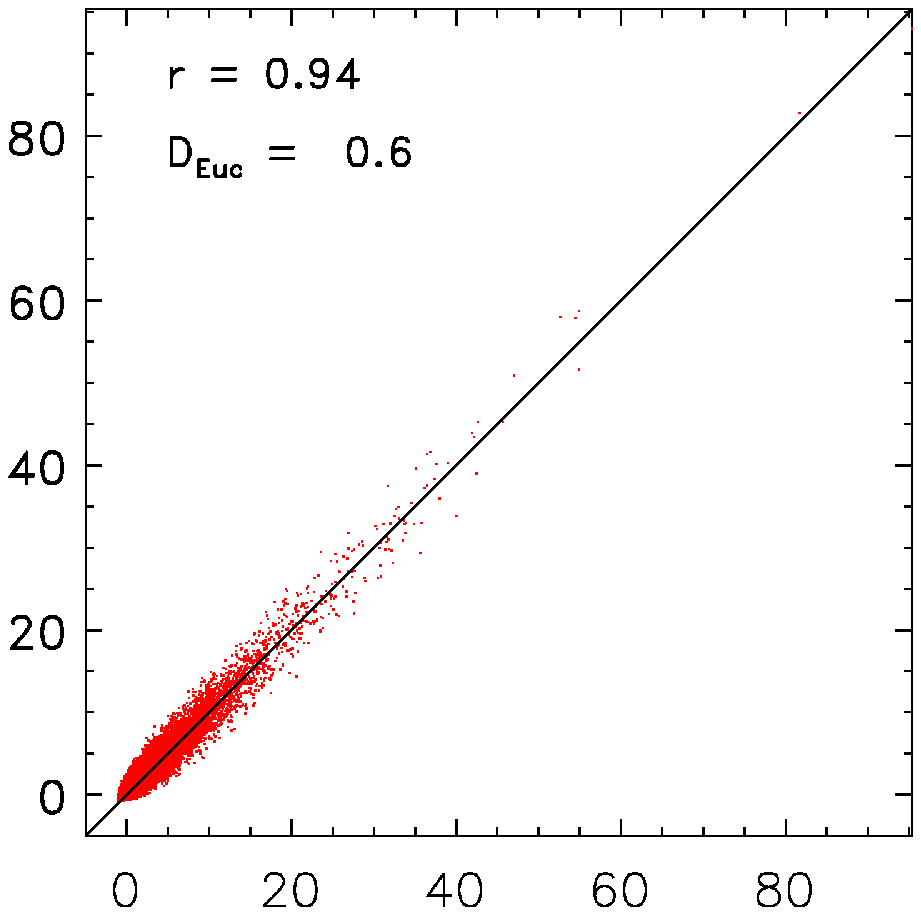}
\put(-185,0.5){{\huge (d)}}
\put(-50,160){\rotatebox[]{0}{\large$64^3$}}
\put(-185,100.){\rotatebox[]{90}{$\delta_{\rm M}^{\rm L}$}}
\put(-80,10){\rotatebox[]{0}{\large$\delta_{\rm M}$}}\hspace{0.5cm}
\end{tabular}
\caption{Matter field reconstructions with the lognormal filter  on a grid mesh with $128^3$ and $64^3$ cells for a uniform selection using about $10^6$ mock galaxies. Panel a: mean over 9 slices through the reconstruction on a mesh with  $128^3$ cells around slice 109 ($Y\sim179$ Mpc/h). Panel b: mean over 5 slices through the reconstruction on a mesh with  $64^3$ cells around slice 55 ($Y\sim179$ Mpc/h). Panels c and d show the cell--to--cell statistics corresponding to the full reconstructions shown in panels a and b, respectively. }
\label{fig:recRES1}
\end{figure*}

\begin{figure*}
\begin{tabular}{cc}
\includegraphics[width=7.cm]{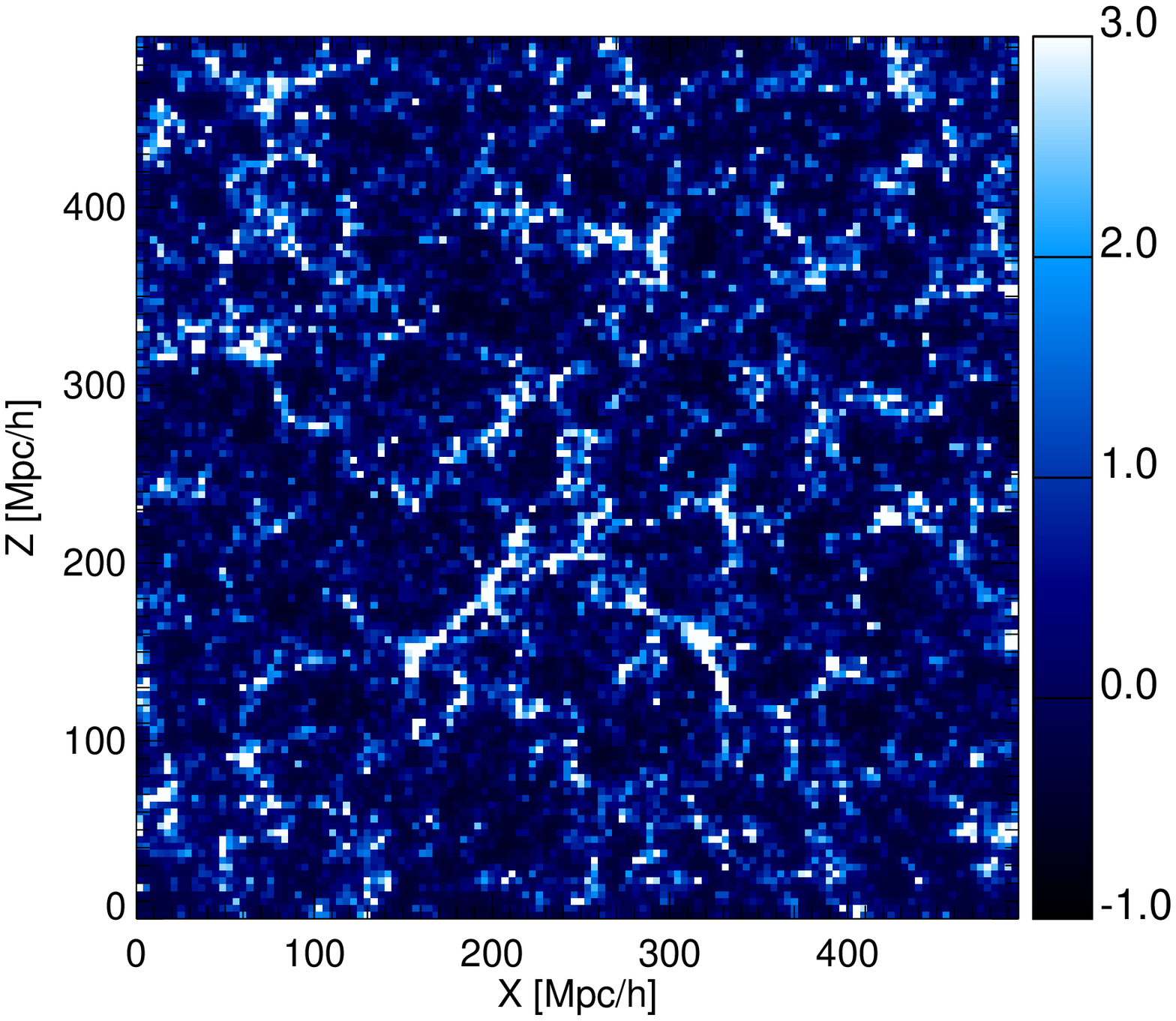}
\put(-190,0.5){{\huge (a)}}
\put(-5,95.){\rotatebox[]{0}{$\delta_{\rm M}^{\rm L}$}}
\hspace{1.5cm}
\includegraphics[width=7.cm]{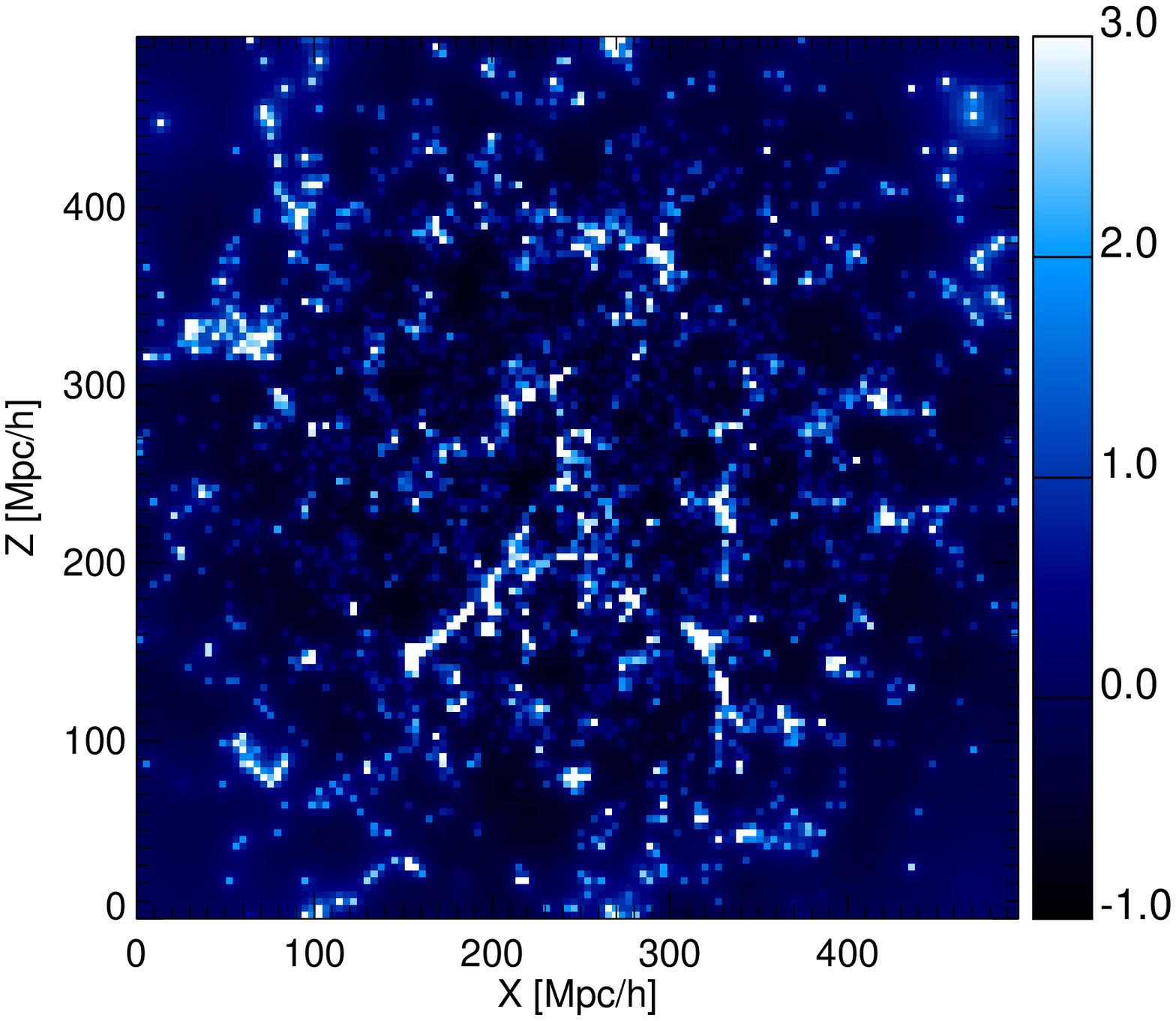}
\put(-190,1.5){{\huge (b)}}
\put(-5,95.){\rotatebox[]{0}{$\delta_{\rm M}^{\rm L}$}}
\vspace{-0.1cm}
\\
\includegraphics[width=5.cm]{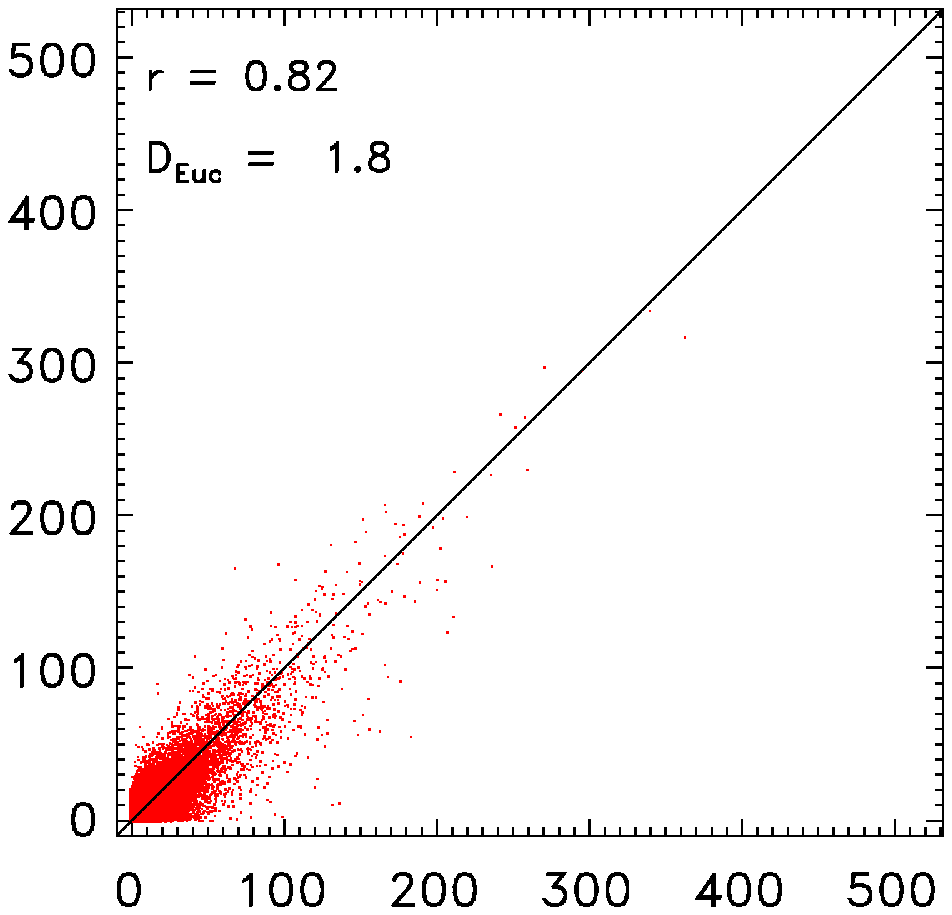}
\put(-160,10){{\huge (c)}}
\put(-50,130){\rotatebox[]{0}{$128^3$}}
\put(-50,140){\rotatebox[]{0}{$w_1$}}
\put(-160,95.){\rotatebox[]{90}{$\delta_{\rm M}^{\rm L}$}}
\put(-70,10){\rotatebox[]{0}{$\delta_{\rm M}$}}


\hspace{3.5cm}
\includegraphics[width=5.cm]{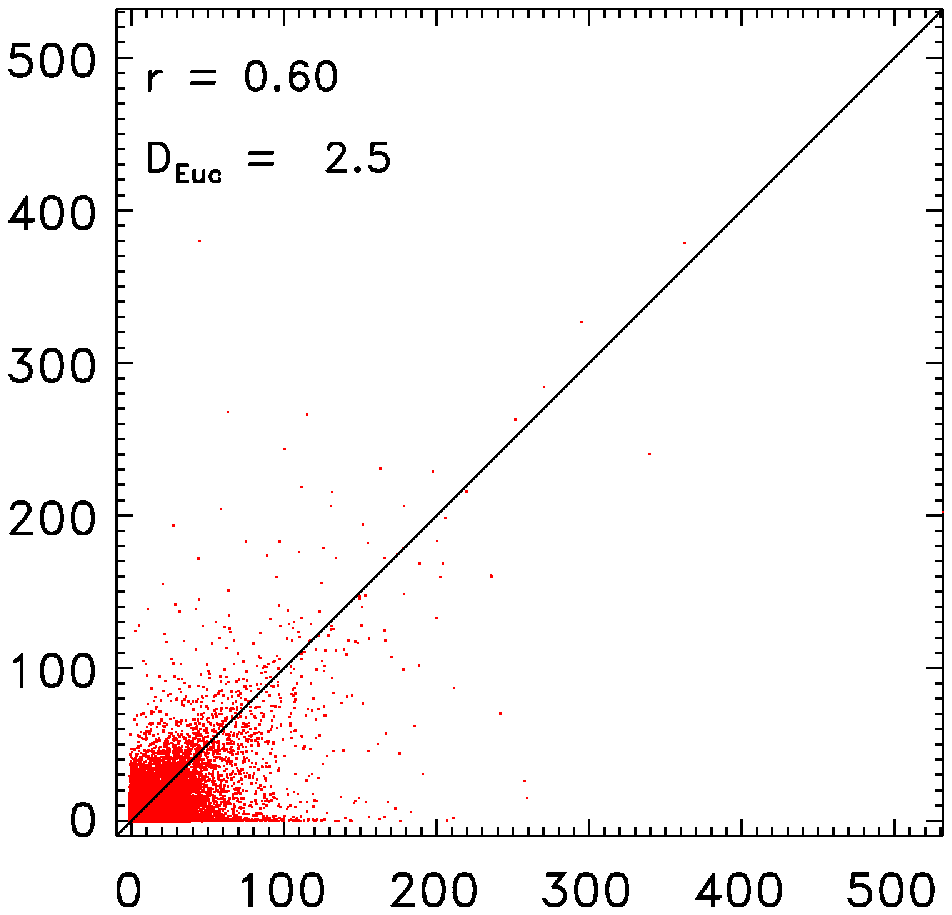}
\put(-160,10){{\huge (d)}}
\put(-50,130){\rotatebox[]{0}{$128^3$}}
\put(-50,140){\rotatebox[]{0}{$w_2$}}
\put(-160,95.){\rotatebox[]{90}{$\delta_{\rm M}^{\rm L}$}}
\put(-70,10){\rotatebox[]{0}{$\delta_{\rm M}$}}

\vspace{-0.3cm}
\\
\includegraphics[width=7.cm]{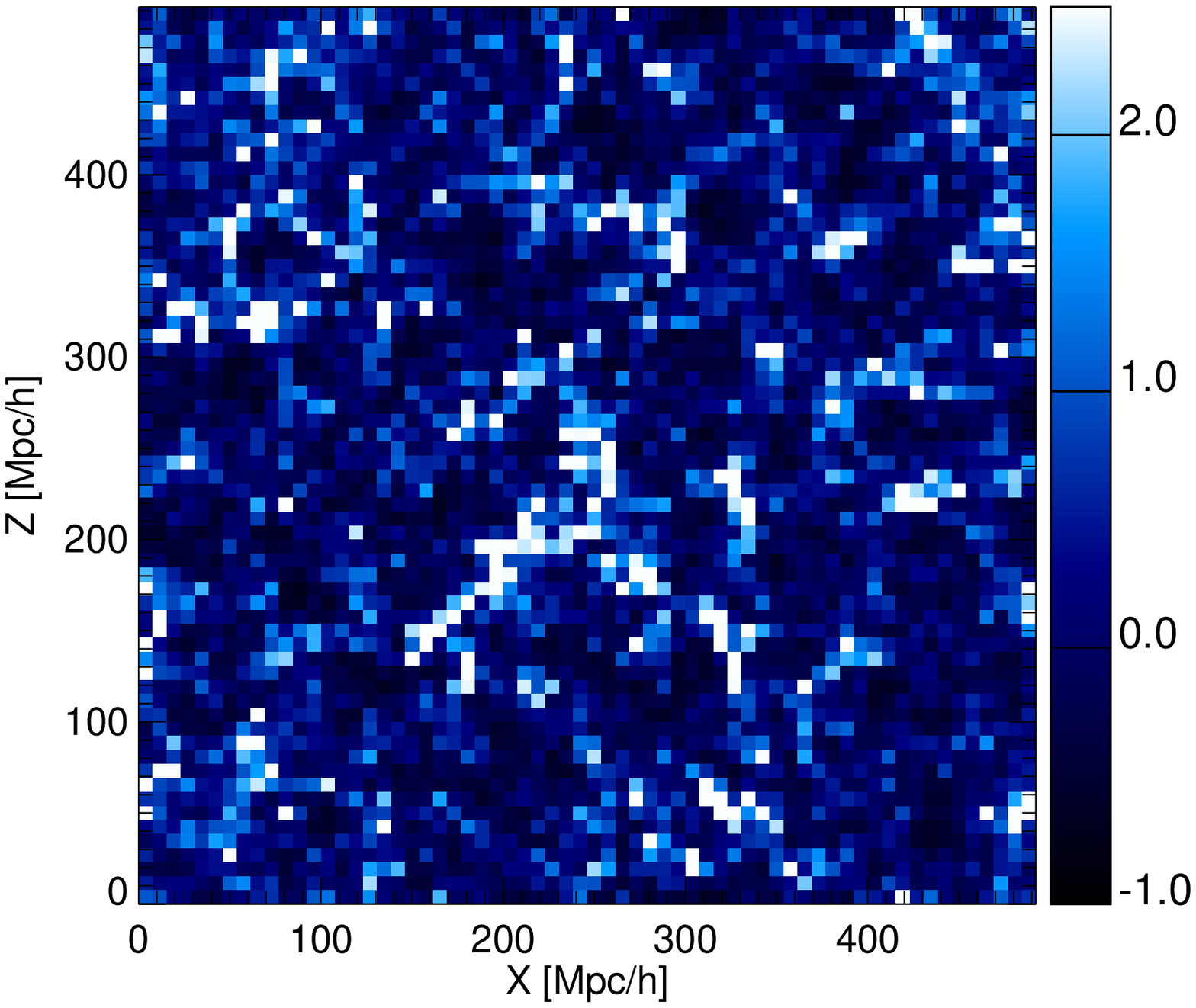}
\put(-190,0.5){{\huge (e)}}
\put(-5,95.){\rotatebox[]{0}{$\delta_{\rm M}^{\rm L}$}}
\hspace{1.5cm}
\includegraphics[width=7.cm]{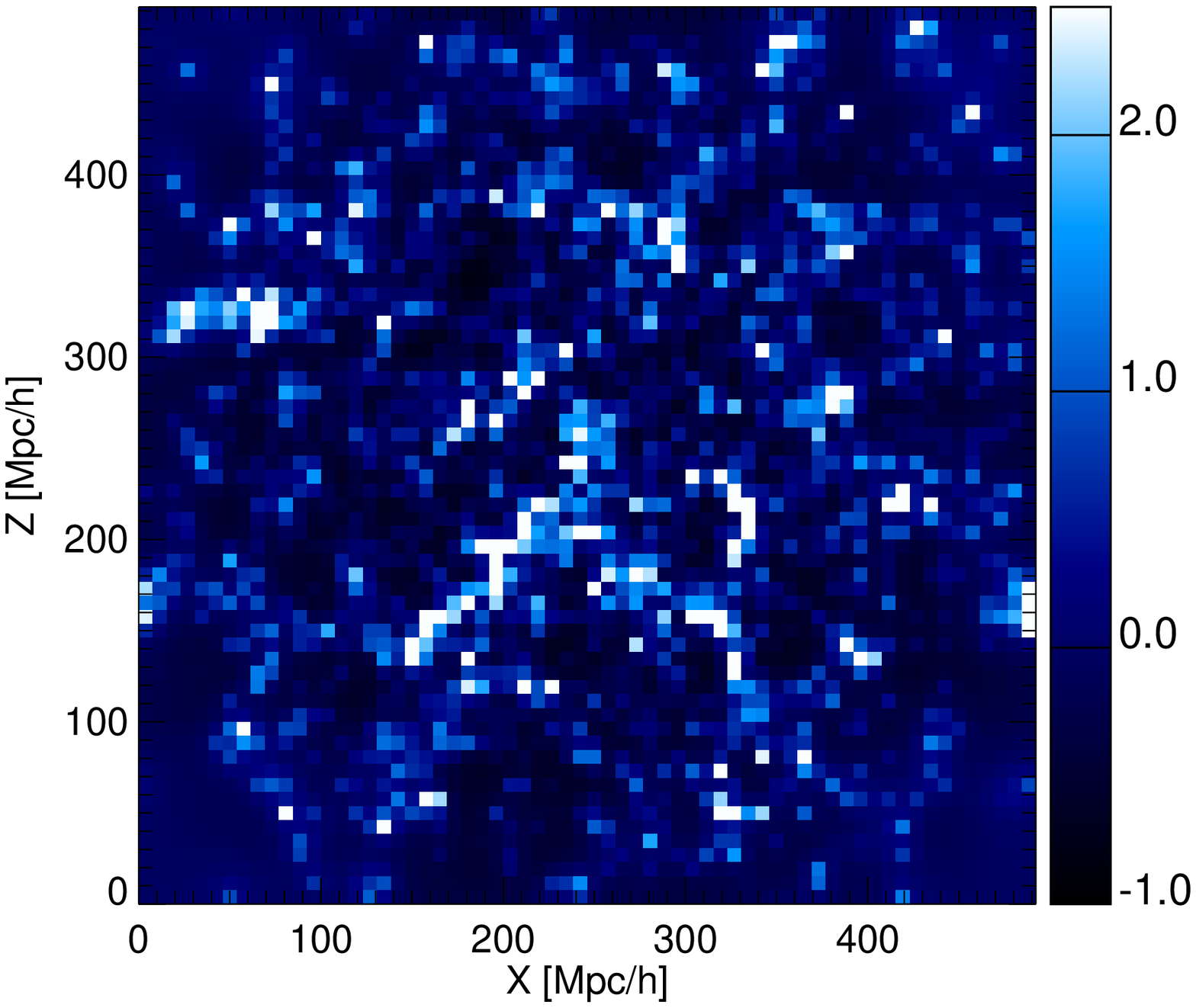}
\put(-190,1.5){{\huge (f)}}
\put(-5,95.){\rotatebox[]{0}{$\delta_{\rm M}^{\rm L}$}}
\\
\includegraphics[width=5.0cm]{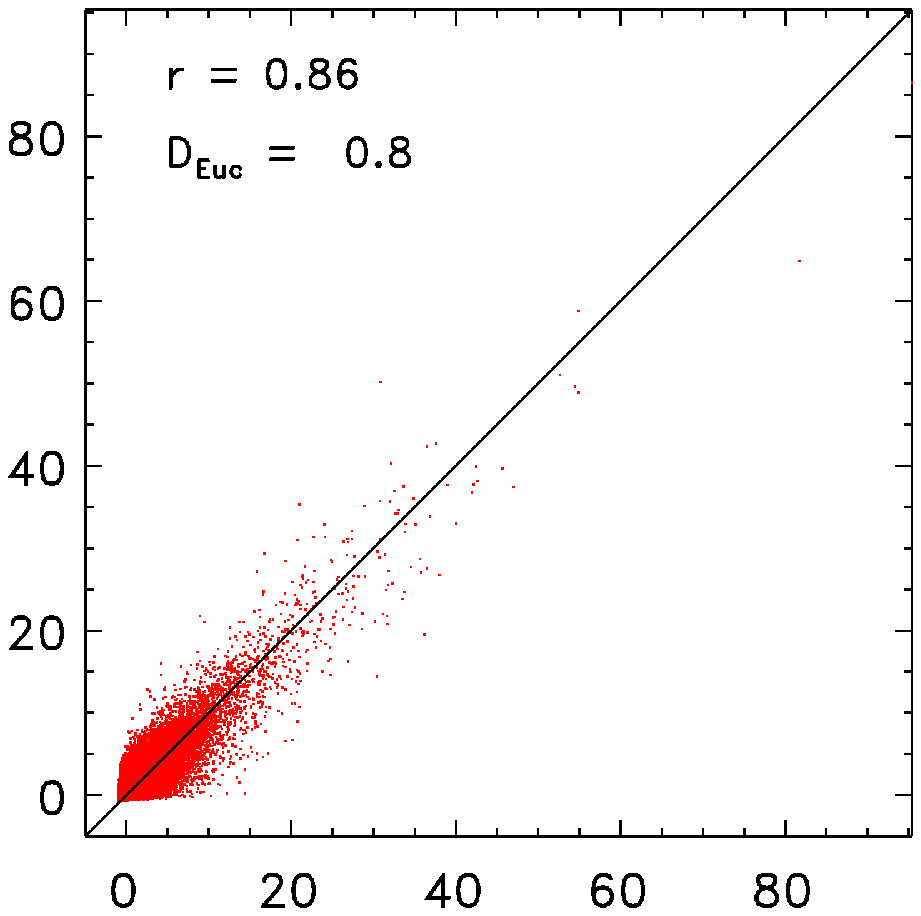}
\put(-160,10){{\huge (g)}}
\put(-50,130){\rotatebox[]{0}{$64^3$}}
\put(-50,140){\rotatebox[]{0}{$w_1$}}
\put(-160,95.){\rotatebox[]{90}{$\delta_{\rm M}^{\rm L}$}}
\put(-70,10){\rotatebox[]{0}{$\delta_{\rm M}$}}

\hspace{3.5cm}
\includegraphics[width=5.0cm]{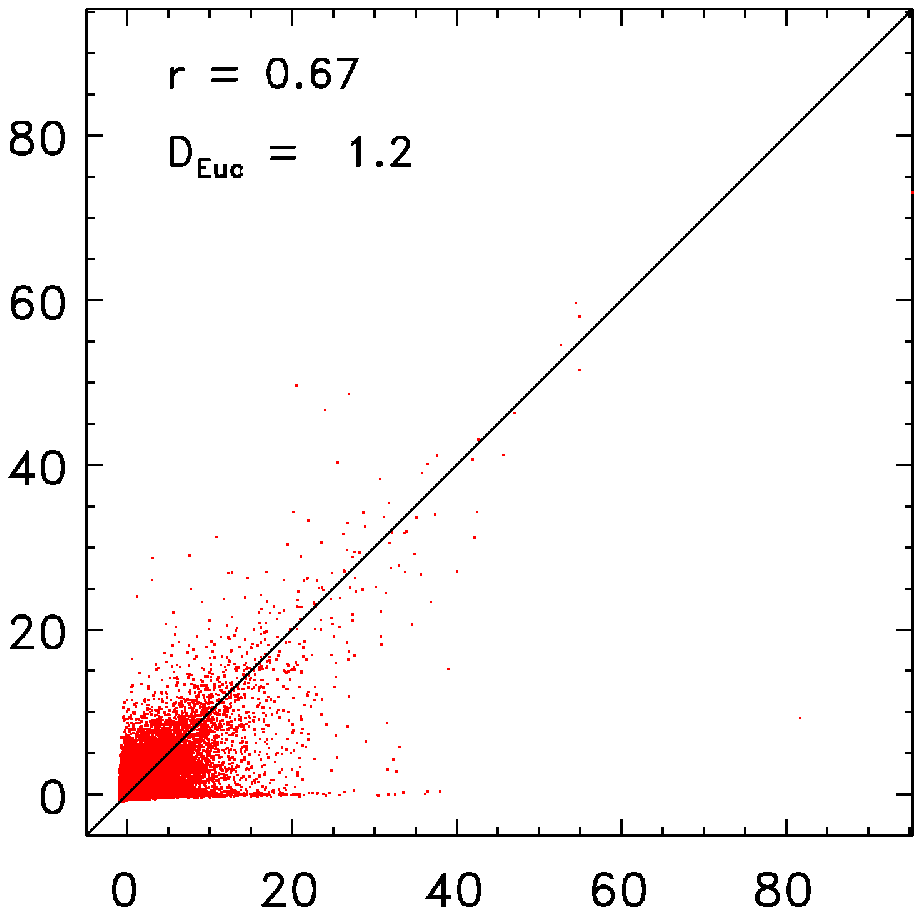}
\put(-160,10){{\huge (h)}}
\put(-50,130){\rotatebox[]{0}{$64^3$}}
\put(-50,140){\rotatebox[]{0}{$w_2$}}
\put(-160,95.){\rotatebox[]{90}{$\delta_{\rm M}^{\rm L}$}}
\put(-70,10){\rotatebox[]{0}{$\delta_{\rm M}$}}

\vspace{-0.5cm}
\end{tabular}
\caption{Matter field reconstructions with the lognormal filter  on a grid mesh with $128^3$ and $64^3$ cells for both $w_1$ and $w_2$ selection criteria. Panel a: same as panel a in previous figure for the case $w_1$. Panel b: same as panel a for the case $w_2$. Panels c and d show the cell--to--cell statistics corresponding to the full reconstructions shown in panels a and b, respectively. Panel e: same as panel a on a mesh with $64^3$. Panel f: same as panel e for the case of $w_2$. Panels g and h show the cell--to--cell statistics corresponding to the full reconstructions shown in panels e and f, respectively.}
\label{fig:recRES}
\end{figure*}

\begin{figure*}
\begin{tabular}{c}
\includegraphics[width=17cm]{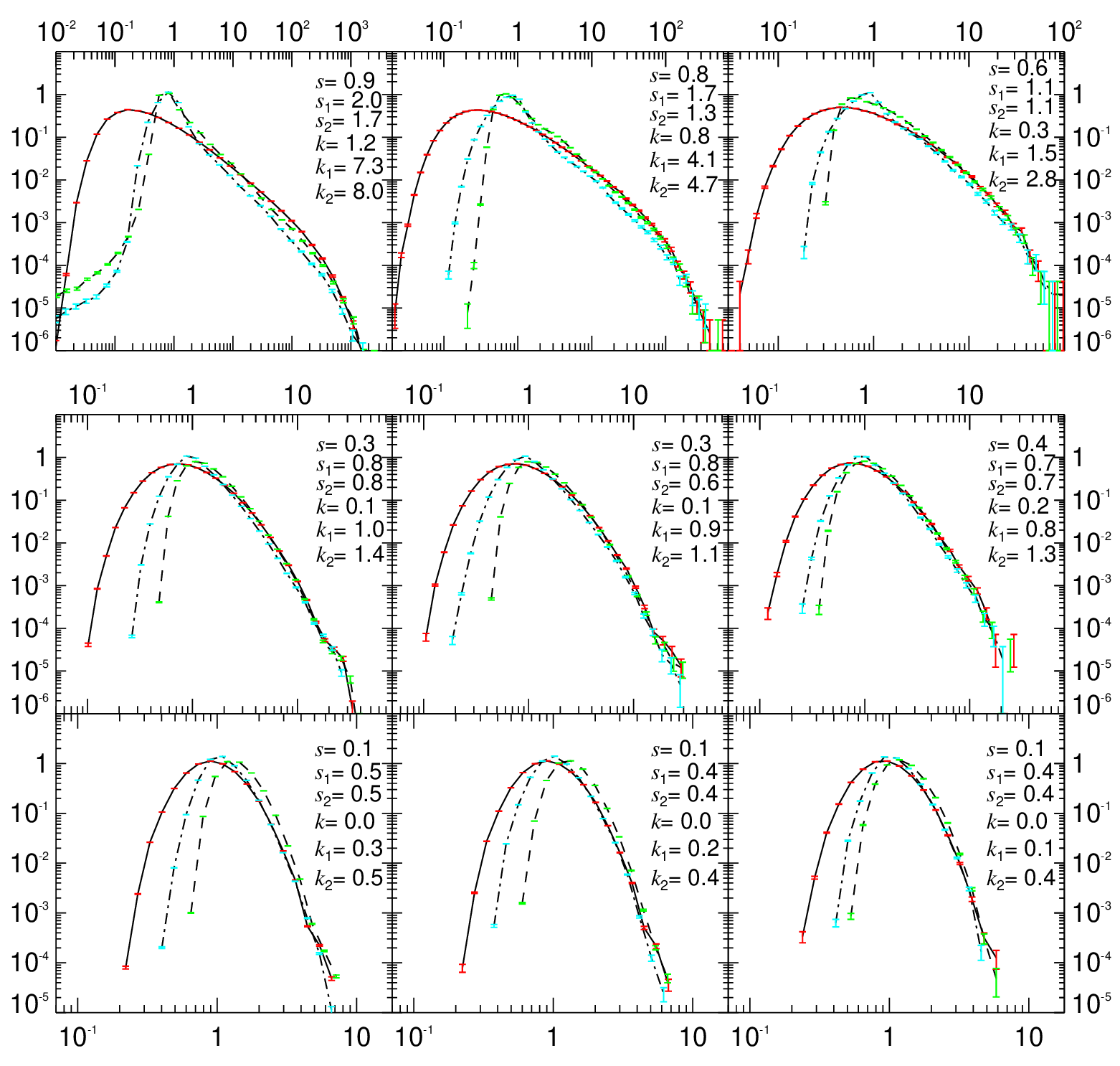}
\put(-400,325){{\huge (a)}}
\put(-260,325){{\huge (b)}}
\put(-100,325){{\huge (c)}}

\put(-400,167){{\huge (d)}}
\put(-260,167){{\huge (e)}}
\put(-100,167){{\huge (f)}}

\put(-400,40){{\huge (g)}}
\put(-260,40){{\huge (h)}}
\put(-100,40){{\huge (i)}}

\put(-495,90){\rotatebox[]{90}{\large$P(\delta_{\rm M})$}}
\put(-495,219){\rotatebox[]{90}{\large$P(\delta_{\rm M})$}}
\put(-495,382){\rotatebox[]{90}{\large$P(\delta_{\rm M})$}}

\put(-100,-3){\rotatebox[]{0}{\large$1+\delta_{\rm M}$}}
\put(-260,-3){\rotatebox[]{0}{\large$1+\delta_{\rm M}$}}
\put(-400,-3){\rotatebox[]{0}{\large$1+\delta_{\rm M}$}}

\put(-160,135){\rotatebox[]{0}{\large$64^3$}}
\put(-160,265){\rotatebox[]{0}{\large$64^3$}}
\put(-160,423){\rotatebox[]{0}{\large$64^3$}}

\put(-160,125){\rotatebox[]{0}{\large 10 Mpc/h}}
\put(-165,255){\rotatebox[]{0}{\large 5 Mpc/h}}
\put(-160,413){\rotatebox[]{0}{\large NGP}}

\put(-305,135){\rotatebox[]{0}{\large$128^3$}}
\put(-305,265){\rotatebox[]{0}{\large$128^3$}}
\put(-305,423){\rotatebox[]{0}{\large$128^3$}}

\put(-305,125){\rotatebox[]{0}{\large 10 Mpc/h}}
\put(-310,255){\rotatebox[]{0}{\large 5 Mpc/h}}
\put(-305,413){\rotatebox[]{0}{\large NGP}}

\put(-450,135){\rotatebox[]{0}{\large$256^3$}}
\put(-450,265){\rotatebox[]{0}{\large$256^3$}}
\put(-450,423){\rotatebox[]{0}{\large$256^3$}}

\put(-450,125){\rotatebox[]{0}{\large 10 Mpc/h}}
\put(-455,255){\rotatebox[]{0}{\large 5 Mpc/h}}
\put(-450,413){\rotatebox[]{0}{\large NGP}}

\end{tabular}
\caption{Matter statistics for the dark matter field from the Millenium run (black curve, red error bars) using about $\sim10^{10}$ particles and the corresponding reconstructions using the selected mocks with radial completeness $w_1$ (dashed curve, green error bars) and $w_2$ (dashed--dotted curve, cyan error bars) having about $\sim10^{5}$ particles for different resolutions ($256^3$: left panels, $128^3$: middle panels  and $64^3$: right panels). Upper panels (a, b and c): without smoothing. Lower panels: after convolution with a Gaussian kernel with smoothing radii of 5 Mpc/h (panels d, e and f) and 10 Mpc/h (panels g, h and i). The number of cells was counted for a logarithmic density binning of 0.2 in $\ln(1+\delta_{\rm M})$ for all cases except for panel a for which a binning of 0.4 was used. Also skewness ($s$, $s_1$ and $s_2$) and kurtosis ($k$, $k_1$ and $k_2$) in $\ln(1+\delta_{\rm M})$  are shown corresponding to the matter field, the reconstructions for the case $w_1$ and the case $w_2$, respectively. The error bars are given by the shot noise caused by the number counts of cells in each density bin without taking into account the uncertainties introduced by the completeness or the reconstruction method itself.} 
\label{fig:stats}
\end{figure*}

We show in Fig.~\ref{fig:rec_MRUS} the performance of the Poisson--lognormal filter with a homogeneous completeness.
Panel a in Fig.~\ref{fig:rec_MRUS} shows a slice through the matter distribution from the Millennium run. Panel b shows the mock galaxy sample. Panels c and d show the Poisson--lognormal filter and the LSQ filter reconstruction respectively. The performance depicted in cell--to--cell correlation plots shown in panels e and f demonstrate the superior behaviour of the Poisson--lognormal filter reconstruction in terms of higher correlation, smaller Euclidean distances and better alignment along the perfect correlation slope. The Poisson--lognormal filter recovers the density field up to overdensities above 1500 whereas the LSQ filter tends to underestimate the density field.

We study the inhomogeneous completeness effects by selecting dark matter particle subsamples with two different radial selection functions depicted in Fig.~\ref{fig:sel}. 
In the upper panel of Fig.~\ref{fig:rec_MRsel}, the inverse weighting scheme is shown to  overestimate  the density at low completeness (at the borders and corners of the cube). This is in agreement with tests performed by \citet[][]{kitaura_sdss}. The LSQ filter, on the other hand, smooths the density more strongly in at low completeness regions and leads to a significantly lower Euclidean distance. The correlation coefficient is lower since the LSQ filtering suppresses the signal and gives a smooth version of the density field which is valid on larger scales (see panels c and d of Fig.~\ref{fig:rec_MRsel}), but does not reproduce small-scale features.  
The lower panels show the results coming from the Poisson--lognormal filter reconstruction. The density at low completeness is suppressed to zero due to the mean field used for this calculation (see Eq.~\ref{eq:mu}). 
In regions of very low completeness the filter tends to favor the mean density.
The statistical correlation shows to be clearly superior to the previous cases and the Euclidean distance with respect to the underlying matter field is far smaller (see panel f). The cell--to--cell correlation plot shows a scatter around the $45^°$ slope and reproduces even the highest overdensities like the one at $\sim$1600 which can also be seen in panel b.
We perform the analogous study with the steeper radial selection function (see Fig.~\ref{fig:sel}). The results are shown in Fig.~\ref{fig:rec_MRsel2} and are consistent with the previously discussed ones.

We perform the same study for two more resolutions: a mesh with $128^3$ cells and a mesh with $64^3$ cells for the same comoving box. First we grid the dark matter field coming from the Millenium run on the lower resolution mesh and then we apply the radial selection $w(r)$ using $w=10^{-4}$, $w(r)=w_1(r)$ and $w(r)=w_2(r)$. The results of the Poisson--lognormal filter reconstructions are shown in Figs.~\ref{fig:recRES1} and \ref{fig:recRES}. We see a clear tendency to better recover the underlying matter field when using a lower resolution (compare Fig.~\ref{fig:rec_MRUS} with Fig.~\ref{fig:recRES1} and Figs.~\ref{fig:rec_MRsel}, \ref{fig:rec_MRsel2} with Fig.~\ref{fig:recRES}).

\subsubsection{Matter statistics}

Finally, we calculate the matter statistics and the corresponding skewness and kurtosis for the dark matter field and the lognormal reconstructions corresponding to the incomplete mocks with selection functions $w_1$ and $w_2$ \citep[for particular expressions to calculate the matter statistics, the skewness and the kurosis see][]{kitaura_sdss}.
The matter statistics represented in Fig.~\ref{fig:stats} shows consistent results for different grid resolutions (compare left, middle and right panels). After convolving the matter fields with a Gaussian kernel using a smoothing radius of 10 Mpc/h the matter distribution appears to be closely lognormal distributed for all resolutions (see panels at the bottom). The skewness ($s$, $s_1$ and $s_2$)  and the kurtosis ($k$, $k_1$ and $k_2$) show some deviation from zero particularly in the reconstructed fields (skewness and kurtosis without subindex correspond to the dark matter field, with the subindex "1" to the selected sample with $w_1$ and with the subindex "2" to the selected sample with $w_2$). However, their values are small which means that the distributions are not especially peaked or have significantly longer tails with respect to the lognormal distribution. This result is consistent with observations \citep[][]{kitaura_sdss} where for a similar smoothing radius the matter field obtained from the Sloan Digital Sky survey (data release 6) was found to be close to lognormal distributed. When convolving with a Gaussian kernel with a 5 Mpc/h  smoothing radius (panels d, e and f) the distribution shows a tail towards larger densities with higher skewness and kurtosis than for the panels at the bottom which cannot be attributed to the uncertainty at the high  densities shown by the large error bars  caused by the low number counts in that regime. This deviation from the  lognormal distribution is even better demonstrated in the upper panels which show the matter statistics without any additional smoothing.
The results show that the multivariate lognormal prior distribution does not impose a lognormal matter field statistics to the recovered density field. Here the prior is subdominant with respect to the data similarly to the case of the LSQ--Wiener reconstruction which can lead to non--Gaussian statistics even though it is based on a Gaussian prior \citep[see][]{kitaura_sdss}.

However, we also observe several effects causing a deviation in the reconstructed fields from the {\it true} matter field statistics.
The reconstructions tend to overestimate the number of cells around the mean density (see the peaks of the distribution in the upper panels of Fig.~\ref{fig:stats}). This trend is more acute for the stronger sampled mock for which $w_2$ was  used. This can be seen by comparing the dashed--dotted curves with the dashed curves and the kurtosis $k_1$ with $k_2$ ($k_2$ is always larger than $k_1$). 
We also observe that the reconstructions underestimate the number of cells in the extremely underdense regions ($\delta\lsim1-0.6$). In addition, we investigated the statistics of the reconstructed fields based on the homogeneously sampled mock data using $w=10^{-4}$ and found that the densities are distributed in a very similar way to the reconstructed matter fields with $w_1$. 
These effects are caused by the conservative character of the reconstruction method. The maximum a posteriori solution leads to a stronger smoothing in the  undersampled low--density regions and produces a larger number of cells with densities closer to the mean.

\section{Conclusions}

In this work we have presented a general expression for the Poisson--lognormal filter given an arbitrary nonlinear galaxy bias. We derived this filter as the maximum a posteriori solution  assuming a lognormal prior distribution for the matter field with a constant mean field and modeling the observed galaxy distribution by a Poissonian process (see Eq.~\ref{eq:MAPLN}). 

We have performed a three--dimensional implementation of this filter  with a very efficient Newton--Krylov inversion scheme (see section \ref{sec:numerics}). Furthermore, we have tested it for a linear galaxy bias relation and compared the results  with other  density field estimators commonly used in the literature (e.g.~the inverse weighting scheme and the least squares (LSQ) Wiener filter (see section \ref{sec:results})).

We also found that the solution of Eqn.~\ref{eq:MAPG}, assuming a Gaussian prior distribution for the matter field, leads to a reconstruction which clearly underestimates large overdensities (see Fig.~\ref{fig:gapmap}). This shows that the Gaussian prior cannot fit the underlying matter field which has a clearly non--Gaussian distribution with a minimum overdensity of $\delta\sim-1$ up to maximal overdensities of about $\delta\sim1700$ for a resolution of $\sim2$ Mpc/h. The density peaks are highly suppressed with the Gaussian prior. This effect is known from the  Wiener filter as traditionally applied in which the noise covariance is dependent on the signal \citep[see discussion in][]{kitaura_sdss}. 


However, we have seen that even the LSQ--Wiener filter fails for high overdensities ($\delta_{\rm M}\gsim1100$).  We showed in appendix \ref{app:LSQ} that the LSQ--filter is the optimal linear filter under a Poisson noise assumption and does not neglect any signal to noise correlation, contrary to what has been assumed in literature \citep[see for example][]{1995ApJ...449..446Z,1998ApJ...503..492S,2004MNRAS.352..939E,kitaura}.  The LSQ filter is the optimal linear filter only up to second order statistics and thus is less well suited to distributions with high skewness or long tails such as the lognormal distribution. Another reason for the inferior performance of the LSQ--filter with respect to the Poisson--lognormal filter is its linearity. Note, that the relation at overdensities $\delta\gg1$ is highly nonlinear.

The one--dimensional lognormal probability distribution is known to fit well the matter distribution up to overdensities of about $\delta\sim$100 as found by \citet{2001ApJ...561...22K}. Our results show, however, good agreement for overdensities even above $\delta\sim$1000 which exceeds by one order of magnitude the regime in which the lognormal is expected to be valid. 
This is because in our filter the lognormal assumption enters as a prior distribution function, but the maximum a posteriori solution is also conditioned on the data. In a similar way \citet[][]{kitaura_sdss} was able to recover a highly non--Gaussian distributed matter field from the SDSS dr6  after using LSQ--Wiener filtering  which according to the Bayesian formalism  assumes a Gaussian distribution. Assuming that the galaxy bias is known we find that the Poisson--lognormal filter is able to recover the matter density fields down to scales of about $\gsim1$2 Mpc/h. However, our study of the  matter statistics comparing the dark matter with the reconstructed fields shows that the Poisson--lognormal filter fails to recover underdense regions for $\delta\lsim1-0.6$.  At lower densities the recovered field is smoothed out due to the conservative maximum a posteriori solution.

Our work shows a great improvement with respect to previous filters in recovering the matter density field from a point source distribution. Still much work has to be done to further analyse the statistical properties of the cosmological structure. Nevertheless, the nonlinear reconstruction method derived in this work could be of great interest for large scale structure density field reconstructions taking a galaxy distribution or even some other observables like the Lyman alpha forest.

\section*{Acknowledgements}

We thank Ofer Lahav and Benjamin D.~Wandelt for suggesting as several
years ago to study the lognormal filter. Special thanks to Rien van de Weijgaert
 and Bernard J.~T.~Jones for discussions about the lognormal prior at
the conference in Santander 2007.  We also thank Simon D.~M.~White,
Carlos Hern\'andez Monteagudo, Torsten En$\ss$lin and Gerard Lemson
for encouraging conversations.

The authors  thank the Intra-European Marie Curie fellowship and
the Transregio TR33 Dark Universe, as well as the Munich cluster {\it
  Universe} for supporting this project and both the Max Planck
Institute for Astrophysics in Munich and the Scuola Internazionale
Superiore di Studi Avanzati in Trieste for generously providing the
authors with all the necessary facilities.

We finally thank the German Astrophysical Virtual Observatory (GAVO),
which is supported by a grant from the German Federal Ministry of
Education and Research (BMBF) under contract 05 AC6VHA, for providing
us with mock data.

{\small
\bibliographystyle{mn2e}
\bibliography{lit}
}


\appendix

\section{The LSQ filter}

\label{app:LSQ}

Here we show a derivation of the LSQ filter which does not require a data degradation model with an additive noise term.
Let us adopt here the usual notation for the data: $\mbi d\equiv\mbi\delta^{\rm o}_{{\rm g}}$.
The data vector is accordingly defined by:
\begin{equation}
\mbi d\equiv\frac{\mbi N^{\rm o}_{\rm g}}{\overline{N}_{\rm g}}-\mat W\vec{1}{,}
\end{equation}
 for a definition of the mask operator $\mat W$ see section \ref{sec:filters}.
We define here the signal vector $\mbi s$ as the matter overdensity field: $\mbi s\equiv \mbi \delta_{\rm M}$.
In the linear approximation we try to find a filter $\mat F$ which applied to the data $\mbi d$ gives an estimate of the signal $\mbi s$ of the form:
\begin{equation}
\langle \mbi s\rangle_{\rm LSQ}\equiv \mat F\mbi d{.}
\end{equation}
This filter should minimize the following quantity in the least squares approach \citep[see][]{1949wiener,1992ApJ...398..169R,1995ApJ...449..446Z}:
\begin{eqnarray}
\cal A&\equiv&\langle \left(\mat F\mbi d-\mbi s\right)^2\rangle\nonumber\\
&=&\mat F\langle \mbi d\mbi d^\dagger\rangle\mat F^{\dagger}-\mat F\langle \mbi d\mbi s^{\dagger}\rangle-\langle \mbi s \mbi d^{\dagger}\rangle\mat F^{\dagger}+\langle \mbi s\mbi s^\dagger\rangle{.}
\end{eqnarray}

As \citet{kitaura} pointed out it is important to note that the ensemble average $\langle \{ \, \} \rangle$ goes over the galaxy and matter field realizations and the filter is thus different from the Wiener filter as derived in a Bayesian framework.  
We define here the global ensemble average by: $\langle \{ \, \} \rangle=\langle\langle \{ \, \} \rangle_{\rm g}\rangle_{\rm M}$. 
 Here $\langle \{ \, \} \rangle_{\rm g} \equiv \langle \{ \, \} \rangle_{(N^{\rm o}_{\rm g}\mid \lambda^{\rm o})} \equiv \sum^{\infty}_{N^{\rm o}_{\rm g}=0} \, P_{\rm Pois}(N^{\rm o}_{\rm g}\mid w\lambda) \{ \, \}$ denotes an ensemble average over the
Poissonian distribution with the expected number of galaxy counts given by the Poissonian
ensemble average: $\lambda^{\rm o}\equiv w\lambda\equiv\langle N^{\rm o}_{\rm g}\rangle_{\rm g}$, and $\langle\{ \, \} \rangle_{\rm M}\equiv\langle\{ \, \} \rangle_{(\delta_{\rm M}\mid \mbi p_{\rm M})}\equiv \int {\rm d}\delta_{\rm M}P(\delta_{\rm M}\mid \mbi p_{\rm M})$ being the ensemble average over all
possible matter density realizations with some prior distribution $P(\delta_{\rm M}\mid \mbi p_{\rm M})$ with $\mbi p_{\rm M}$ being a set of parameters which determine the matter field, say the cosmological parameters.
We impose $\langle\delta_{\rm M}\rangle_{\rm M}=0$.

Recalling the derivations done by \citet[][]{1949wiener,1992ApJ...398..169R,1995ApJ...449..446Z} we find minimizing the action with respect to the filter:
\begin{equation}
\frac{\partial \cal A}{\partial \mat F}=0{,}
\end{equation}
the following LSQ filter expression:
\begin{equation}
\label{eq:FLSQ}
\mat F=\langle \mbi s \mbi d^\dagger\rangle\langle \mbi d\mbi d^\dagger\rangle^{-1}{.}
\end{equation}
Traditionally one would then define a data degradation model with an additive noise term of the form: $\mbi d=\mat R\mbi s+\mbi \epsilon$, with $\mat R$ being some response operator. Then substituting this data model in Eq.~\ref{eq:FLSQ} and neglecting noise to signal correlation terms one would obtain a final expression for the LSQ filter \citep[see][]{1995ApJ...449..446Z}.

\subsection{Signal to noise correlation}
One can show that the noise is actually uncorrelated with the signal by making the following definition:
\begin{equation}
\epsilon^{\rm o}_i\equiv N^{\rm o}_{{\rm g},i}-\langle N^{\rm o}_{{\rm g},i}\rangle_{\rm g}{,}
\label{app:noise}
\end{equation} 
and then calculating the correlation:
\begin{equation}
\langle \epsilon^{\rm o}_i\langle N^{\rm o}_{{\rm g},j}\rangle_{\rm g}\rangle_{\rm g}= \langle N^{\rm o}_{{\rm g},i}\langle N^{\rm o}_{{\rm g},j}\rangle_{\rm g}-\langle N^{\rm o}_{{\rm g},i}\rangle_{\rm g}\langle N^{\rm o}_{{\rm g},j}\rangle_{\rm g}\rangle_{\rm g}=0{.}
\end{equation} 
Note, that this implies: $\langle \epsilon^{\rm o}_i\langle \delta^{\rm o}_{{\rm g},j}\rangle_{\rm g}\rangle_{\rm g}=0$ and thus also $\langle \mbi \epsilon\mbi s^\dagger\rangle_{\rm g}=0$.

\subsection{LSQ filter derivation without the additive noise assumption}

However, one does not even need to use the additive noise assumption to derive the LSQ filter. Let us show here how to make such a derivation.
We define the observed galaxy number counts per cell $i$ as:
\begin{equation}
N^{\rm o}_{{\rm g},i}\equiv\overline{N}_{{\rm g}}(w_i+\delta_{{\rm g},i}^{\rm o}){.}
\end{equation}
The corresponding ensemble average over all possible galaxy realizations is:
\begin{equation}
\langle N^{\rm o}_{{\rm g},i}\rangle_{\rm g}\equiv\overline{N}_{{\rm g}} w_i(1+\delta_{{\rm g},i}){.}
\label{app:binomial}
\end{equation}
Recalling the linear bias relation:
\begin{equation}
\delta_{{\rm g},i}=\sum_jb_{i,j}\delta_j{,}
\end{equation} 
we can then calculate with the above definitions the signal to data correlation matrix: 
\begin{eqnarray}
\langle s_{i}d_{j}\rangle&\equiv&\langle\delta_{{\rm M},i}\delta^{\rm o}_{{\rm g},j}\rangle = \langle\langle\delta_{{\rm M},i}\delta^{\rm o}_{{\rm g},j}\rangle_{\rm g}\rangle_{\rm M}\nonumber\\
&=&\langle\delta_{{\rm M},i}\langle \delta^{\rm o}_{{\rm g},j}\rangle_{\rm g}\rangle_{\rm M}=w_j\sum_{j'}b_{j,j'}\langle\delta_{{\rm M},i}\delta_{{\rm M},j'}\rangle_{\rm M}{.}
\end{eqnarray}
We also have to calculate the data autocorrelation matrix:
\begin{eqnarray}
\langle d_{i}d_{j}\rangle\equiv\langle\delta^{\rm o}_{{\rm g},i}\delta^{\rm o}_{{\rm g},j}\rangle&=&\langle\langle \left(\frac{N^{\rm o}_{{\rm g},i}}{\overline{N}_{{\rm g}}}-w_i\right)\left(\frac{N^{\rm o}_{{\rm g},j}}{\overline{N}_{{\rm g}}}-w_j\right)  \rangle_{\rm g}\rangle_{\rm M}\nonumber\\
&=&\frac{\langle\langle N^{\rm o}_{{\rm g},i}N^{\rm o}_{{\rm g},j}\rangle_{\rm g}\rangle_{\rm M}}{\overline{N}_{{\rm g}}^2}-w_iw_j{.}
\end{eqnarray}
Here we need a model for the two-point number count statistics. Note, that we can introduce here Poissonity:
\begin{equation}
\langle N^{\rm o}_{{\rm g},i}N^{\rm o}_{{\rm g},j}\rangle_{\rm g}\equiv\langle N^{\rm o}_{{\rm g},i}\rangle_{\rm g}\langle N^{\rm o}_{{\rm g},j}\rangle_{\rm g}+\langle N^{\rm o}_{{\rm g},i}\rangle_{\rm g}\delta^{\rm K}_{i,j}{.}
\end{equation}
With the additional matter field ensemble average we get:
\begin{equation}
\langle\langle N^{\rm o}_{{\rm g},i}\rangle_{\rm g}\langle N^{\rm o}_{{\rm g},j}\rangle_{\rm g}\rangle_{\rm M}=\overline{N}^2_{{\rm g}}w_iw_j\left(1+\sum_{k}b_{i,k}\sum_{l}b_{j,l}\langle \delta_{{\rm M},k}\delta_{{\rm M},l}\rangle_{\delta}\right){.}
\end{equation}
We can define the noise covariance matrix as:
\begin{eqnarray}
N_{i,j}&\equiv&\frac{1}{\overline{N}^2_{{\rm g}}}\langle\langle N^{\rm o}_{{\rm g},i}N^{\rm o}_{{\rm g},j}\rangle_{\rm g}-\langle N^{\rm o}_{{\rm g},i}\rangle_{\rm g}\langle N^{\rm o}_{{\rm g},j}\rangle_{\rm g}\rangle_{\rm M}\nonumber \\
&=&\frac{1}{\overline{N}^2_{{\rm g}}}\langle\langle N^{\rm o}_{{\rm g},i}\rangle_{\rm g}\rangle_{\rm M}\delta^{\rm K}_{i,j}=\frac{w_i}{\overline{N}_{{\rm g}}}\delta^{\rm K}_{i,j}{.}
\end{eqnarray}
The LSQ filter can be then written as:
\begin{eqnarray}
\lefteqn{F_{i,j}=\sum_{j'}w_{j'}\sum_{l}b_{j',l}\langle \delta_{{\rm M},i}\delta_{{\rm M},l}\rangle_{\rm M}}\\
&&\times\left( w_{j'}\sum_{k}b_{j',k}\sum_{k'}b_{j,k'}\langle \delta_{{\rm M},k}\delta_{{\rm M},k'}\rangle_{\rm M}w_j+\frac{w_j}{\overline{N}_{{\rm g}}}\delta^{\rm K}_{{j'},j}\right)^{-1}\nonumber{.}
\end{eqnarray}
The corresponding matrix notation of the LSQ filter yields:
\begin{equation}
\mat F=\mat S\mat R^\dagger\left(\mat R\mat S\mat R^\dagger+\mat N\right)^{-1}{,}
\end{equation}
with $\mat S\equiv\langle \mbi\delta_{{\rm M}}\mbi\delta_{{\rm M}}^\dagger\rangle_{\rm M}$ and $\mat R\equiv \mat W\mat B$ (see section \ref{sec:filters} for a definition of the bias operator $\mat B$).
 This data-space expression is equivalent to the signal-space representation   \citep[for a demonstration see appendix C in][]{kitaura}:
\begin{equation}
\mat F=\left(\mat S^{-1}+\mat R^\dagger\mat N^{-1}\mat R\right)^{-1}\mat R^\dagger\mat N^{-1}{.}
\end{equation}
We conclude that the LSQ filter is the optimal linear filter under a Poisson noise assumption. We have shown that this filter does not neglect any signal to noise correlation.

\section{Lognormal prior and Gaussian likelihood}

\label{app:LPGL}

For completeness we derive a nonlinear filter which assumes a lognormal prior and a Gaussian likelihood. Following \citet[][]{1995MNRAS.277..933S} one could use the Wiener filter with a data transformation and apply it to recover non-Gaussian distributed fields. The problem in such a model is that one requires a multiplicative noise assumption of the form: 
\begin{eqnarray}
d'_i&\equiv&\delta_i\epsilon'_i+\epsilon'_i\nonumber \\
\ln(d'_i)&=&\ln(1+\delta_i)+\ln(\epsilon'_i)\nonumber \\
d_i&\equiv&s_i+\epsilon_i{,}
\end{eqnarray}
for each cell $i$, with $d_i\equiv\ln(d'_i)$, $s_i\equiv\ln(1+\delta_i)$ and $\epsilon_i\equiv\ln(\epsilon'_i)$.
Note, that with such a data model one could easily apply the Wiener filter assuming that the signal $s$ and the noise $\epsilon$ are Gaussian distributed.

\subsection {Additive noise model}

However, one may rather prefer a data model with an additive noise term as commonly used in the literature \citep[see e.g.~][]{1995ApJ...449..446Z,1997ApJ...480L..87T}.
We define therefore a data model of the form:
\begin{equation}
\mbi d\equiv \mat R\mbi \delta+\mbi \epsilon{,} 
\end{equation}
including in the signal higher order terms: 
\begin{equation}
\mbi \delta=\exp(\mbi s)-\vec{1}{.}
\label{eq:expdata}
\end{equation}
One can then assume the signal to be lognormal distributed and the noise to be Gaussian distributed and signal-independent.

\subsection{Gaussian likelihood}

Let us write the log-likelihood as:
\begin{equation}
\ln{\mathcal{L}}\propto -\frac{1}{2}\left(\mbi \epsilon^\dagger\mat N^{-1}\mbi \epsilon-\ln\left(\det\left(\mat N\right)\right)\right)+c{.}
\end{equation}
Making the substitution $\mbi\epsilon=\mbi d-\mat R\mbi \delta$ we get:
\begin{equation}
\mbi \epsilon^\dagger\mat N^{-1}\mbi \epsilon=\mbi d^\dagger\mat N^{-1}\mbi d+\mbi \delta^{\dagger}\mat R^\dagger\mat N^{-1}\mat R\mbi \delta-\mbi \delta^{\dagger}\mat R^\dagger\mat N^{-1}-\mbi d^\dagger \mat N^{-1}\mat R\mbi \delta{.}
\end{equation}
To find the maximum a posteriori solution we have to calculate the derivative of the likelihood with respect to the signal $\mbi s$:
\begin{equation}
\sum_i\frac{\partial\ln {\cal L}_i}{\partial s_k}=\sum_i\sum_l\frac{\partial\ln {\cal L}_i}{\partial \delta_l}\frac{\partial \delta_l}{\partial s_k}{.}
\end{equation} 
From Eq.~\ref{eq:expdata} we get:
\begin{equation}
\frac{\partial \delta_l}{\partial s_k}=\exp(\mbi s_i)\delta^{\rm K}_{i,k}{.}
\end{equation} 
Assuming a signal independent noise yields:
\begin{equation}
\frac{\partial\ln {\cal L}_i}{\partial \delta_k}= -\sum_{jlm} \delta_jR_{j,l}N^{-1}_{l,m}R_{m,i}+\sum_{jl}d_jN_{j,l}^{-1}R_{l,i} {.}
\end{equation} 
Combining these results with the derivative of the lognormal prior (Eq.~\ref{eq:priorlog}) leads to:
\begin{eqnarray}
\lefteqn{\sum_jS^{-1}_{{\rm L}j,k}\left(s_j-\mu_j\right)=}\\
&&\left(\sum_{jlm}\left(\exp(s_j)-\vec{1}\right)R_{j,l}N^{-1}_{l,m}R_{m,k}+\sum_{jl}d_jN^{-1}_{j,l}R_{l,k}\right)\exp(s_k)\nonumber{.}
\end{eqnarray}

\end{document}